\documentclass{emulateapj}
\usepackage{psfig}

\def\mincir{\ \raise -2.truept\hbox{\rlap{\hbox{$\sim$}}\raise5.truept  
\hbox{$<$}\ }}                        % minore o circa uguale
\def\magcir{\ \raise -2.truept\hbox{\rlap{\hbox{$\sim$}}\raise5.truept  
\hbox{$>$}\ }}                        % maggiore o circa uguale

\slugcomment{Draft version}

\shorttitle{Composite Seyfert/Star-forming galaxies}
\shortauthors{}

\begin{document}

\title{The nature of Composite Seyfert/Star-forming galaxies revealed
by X-ray observations}

\author{Francesca Panessa\altaffilmark{1,2}, Anna
   Wolter\altaffilmark{3}, Silvia Pellegrini\altaffilmark{4}, 
   Antonella Fruscione\altaffilmark{2},
Loredana Bassani\altaffilmark{5}, Roberto Della Ceca\altaffilmark{3},
Giorgio G.C. Palumbo\altaffilmark{4}, Ginevra Trinchieri\altaffilmark{3}}

\altaffiltext{1}{Instituto de Fisica de Cantabria (CSIC-UC),
Avda. de los Castros, 39005 Santander, Spain }
\altaffiltext{2}{Center for Astrophysics, 60 Garden 
Street, Cambridge, MA 02138}
\altaffiltext{3}{Osservatorio Astronomico di Brera,
Via Brera 28, 20121 Milano, Italy}
\altaffiltext{4}{Dipartimento di Astronomia, Universit\`a di Bologna,
via Ranzani 1, 40127 Bologna, Italy}
\altaffiltext{5}{Istituto di Astrofisica Spaziale e Fisica Cosmica (IASF-CNR), 
Via Gobetti 101, 40129 Bologna, Italy} 

\begin{abstract}

This paper presents new {\it Chandra} and {\it Beppo}SAX 
observations aimed at investigating the
optical/X-ray mismatch in the enigmatic class of the Composite
galaxies, discovered by a 
cross-correlation of IRAS and
ROSAT all sky survey catalogues. These galaxies have been classified as
star-forming objects on the basis of their optical spectra, while the
detection of weak broad wings in the H$\alpha$ emission in a few of them 
and their high
X-ray luminosity in the ROSAT band indicated the presence of an active
nucleus. The analysis of {\it Chandra}  observations for 4 Composites 
has revealed nuclear point-like sources, 
with a typical AGN spectrum ($\Gamma$$\sim$ 1.7-1.9) and
little intrinsic absorption. A strong flux
variability has been observed on different time scales, in particular
most of the sources were brighter at the ROSAT epoch. Although of relative low
luminosity for the AGN class (L$_{2-10 keV}$ $\sim$ 3-60 $\times$ 10$^{41}$
erg/s), the active nucleus is nevertheless dominant in the X-ray domain.
At other wavelengths it appears to be overwhelmed by the starburst 
and/or host galaxy light, yielding the Composite 
classification for these objects.

\end{abstract}


\section{Introduction}

Evidence for a link between intense star formation and nuclear
activity has grown steadily in recent years (e.g. Veilleux, 2001,
Gonzales Delgado et al. 2001, and references therein).  
On one hand,
the presence of circumnuclear starbursts in many local AGNs
(e.g., Levenson et al. 2004, Levenson et al. 2001) suggests a
connection not yet fully understood. The presence of a starburst has been
invoked for instance to produce
absorption in low luminosity AGNs (Fabian et al. 1998, Ohsuga \&
Umemura, 2001).
On the other hand,
the ubiquity of supermassive black holes in the nuclei of normal
galaxies (Kormendy et al. 2002) and the proportionality between the
black hole and the spheroidal masses (Ferrarese \& Merrit 2000)
evidence
a direct link between the formation of
ellipticals and spiral bulges and the growth of central black holes.
Therefore, the interplay between accretion on supermassive black holes
and galaxy formation and
evolution has become a fundamental ingredient for theoretical models
in this field (e.g., Springel et al. 2005, Sazonov et al. 2005, Nipoti
et al. 2003).
This implies that our view on
the star formation history of the universe, as deduced by galaxy
luminosity functions (see Springel et al. 2005), 
as well as on the chemical enrichment and
feedback processes in the early universe, might change in a way that we cannot
foresee at this moment.
Understanding the connection between starburst and AGN in the local
universe is therefore of crucial importance.

A spectroscopic optical survey of bright IRAS and X-ray selected
sources from the ROSAT All Sky Survey revealed a small enigmatic class of
low redshift galaxies with optical spectra dominated by the features
of H~II galaxies but presenting very subtle Seyfert signatures
as well. These objects had
X-ray luminosities typical of broad line AGNs, ranging from
1.5 $\times$ 10$^{42}$ erg/s to 5 $\times$ 10$^{43}$ erg/s in the
ROSAT band (Moran et al. 1996, hereafter M96).  They were named
Composite after their Composite optical spectra.
The diagnostic emission line ratio diagrams (Veilleux
\& Osterbrock 1987) classify them either on the boundary
between Seyfert and HII regions or as pure star-forming
galaxies. Most of these galaxies show quite narrow emission lines
(FWHM$<$300 km/s) as HII galaxies normally do. Yet, some of them
present [O~III]$\lambda$$\lambda$4959,5007 lines significantly broader than all
other narrow lines in the spectrum and weak and elusive broad
H$\alpha$ wings. 
A possible scenario to explain the optical and X-ray mismatch proposed
by M96 invokes absorption as responsible for the
obscuration of the optical features.
M96 also suggested that it is not likely that the
starburst could overpower a Seyfert optical nuclear spectrum,
given that the starburst component in these objects is not particularly
strong. On the contrary here
we propose that the AGN and the starburst components in these objects are 
present at the same level of activity making them 
ideal laboratories in which the two phenomena can be studied together.

Until recently little or no information in the X-ray spectral domain 
were available
for the M96 sample of Composite galaxies except for the ROSAT
data. In the last couple of years,
{\it Chandra} and XMM-{\it Newton} observations of IRAS\,00317$-$2142
and IRAS\,20051$-$1117, two of the brightest objects of the sample,
have revealed the presence of an active nucleus with a spectral shape
typical of type 1 AGN's (a photon index $\Gamma$ $\sim$ 1.8-1.9) and no 
absorption.
Long term variability has been observed in IRAS\,00317$-$2142
(Georgantopoulos, Zezas \& Ward 2003), while no flux variation has been
measured in IRAS\,20051$-$1117 (Georgantopoulos et al. 2004).

In order to acquire the missing information in X-rays for the rest of
the sample, and to derive the overall properties for the whole sample, we
first proposed to observe these objects with {\it Beppo}SAX (Boella et al.
1997) and we
obtained data for 3 sources of the original list of 6 of Moran et
al. (1996). Then we observed four objects of the sample with
{\it Chandra} (Weisskopf et al. 2000).
Here we present the results obtained from the {\it Chandra} and {\it Beppo}SAX
observations and discuss possible explanations
for the enigmatic behaviour of this class of objects.

The paper is organized as follows: the sample is briefly outlined in
Section 2, new X-ray observations
are described in Sections 3 through 6 and discussed together
with a multi-frequency analysis in Section 7.
Our conclusions are drawn in Section 8.
Throughout the paper we assume H$_{0}=75$ km s$^{-1}$ Mpc$^{-1}$.

\section{The Composite sample}

We list the 6 objects defined as Composite by M96
in Table~\ref{sample}. 
We note that the original M96's list included 7 objects,
but IRAS\,10113+1736 is no longer a valid Composite, since it has
become evident that infrared and
X-ray emissions originate from different sources (Condon et al. 1998a).

Table~\ref{sample} lists all the sample properties, i.e. J2000 position, 
redshift, Hubble type (from LEDA\footnote{\tt{http://leda.univ-lyon1.fr/
}}, Paturel et al. 1997),
Galactic N$_{\rm H}$ from Dickey \& Lockman (1990), IRAS far-infrared 
luminosities;
we also report the logarithmic luminosities of the H${\alpha}$
(narrow+broad and only broad) and [O~III]$\lambda$5007
emission lines obtained by M96 
in a $2^{\prime\prime}\times4^{\prime\prime}$ aperture.

We note that IRAS\,00317$-$2142 belong to the compact group HCG 4. However,
no extended emission is detected in the group at a limit of 10$^{41}$ erg
s$^{-1}$ in the ROSAT PSPC (Mulchaey et al. 2003), so that no group emission
is expected to contaminate our results.

The galaxies are local (z $<$ 0.04). They are all detected by the
NVSS (Condon et al. 1998b) in the range 6-44 mJy and by the  2MASS 
(Cutri et al. 2003) with K$_s$ mag between 10.3 and 11.6.
Their optical magnitudes range between 14 and 16
(from NED). 

For completeness we include the results of the ASCA,
XMM-{\it Newton} and {\it Chandra} observations of IRAS\,00317$-$2142 and
IRAS\,20051$-$1117 (Georgantopoulos 2000, Georgantopoulos, Zezas \& Ward
2003, Georgantopoulos et al. 2004) in the X-ray table results (see Sect. 5)
and in the final discussion of the whole sample.

\section{X-ray data}

\subsection{{\it Beppo}SAX Observations}

We observed three Composite objects with the Narrow Field Instruments
(NFI) of the {\it Beppo}SAX satellite. 
The observation dates and the
total effective exposure time in ks for the LECS
(0.1-4 keV), MECS (1.3-10 keV) and PDS (13-200 keV)
instruments are listed in the bottom part of Table~\ref{infobs}.
All objects have been detected by the MECS instrument, none
by the LECS detector.
The only source detected by the PDS is IRAS\,20051$-$1117
(S/N $\magcir$ 3) with a total
count rate of (13.73$\pm$4.48)$\times$10$^{-2}$ cts/s in the 20-200 keV band;
  however the PDS data fall above the simple power-law
extrapolation from the MECS data suggesting
a possible contamination from a different object
as discussed in Sec 4.

The extraction of the
source and background spectra was done with the XSELECT package.
The MECS spectra have been extracted from a 4 arcmin radii circular
regions; the background for each source has been  extracted from source
free off-axis circular regions and subtracted. 
The redistribution matrices and
ancillary response files released in September 1997 
have been applied. MECS net count rates and errors
are listed in Table~\ref{infobs}.

\subsection{Chandra Observations}

We observed four objects with the Advanced CCD Imaging
Spectrometer ACIS-I on board {\it Chandra} between March and September
2003 using exposures of $\sim$ 25 ks. Details of the observations
can be found in the top part of Table~\ref{infobs}. The data reduction was
performed using the CIAO version 3.1 software. Level 1 event files
have been created applying the time-dependent gain correction.  The
standard procedure provided by the CIAO ``threads" has been followed in
order to obtain level 2 event files.  Background light curves have
been examined and in two cases (IRAS\,01072+4954 and IRAS\,01319$-$1604)
background flares have been found and removed. In all cases, ``good''
exposure times (listed in column 3 of Table~\ref{infobs})
are always $\geq$ 90\% of the total observation times.
Based on the RASS
fluxes, we had requested observations in 1/4 sub-array mode in order to
minimize pile-up effects. However, we found that the observed count rates
are below the expected values, therefore no pile-up affects the 
observations.

Nuclear spectra have been extracted from a circular region of a 
3$^{\prime\prime}$ radius; 
  the background has been evaluated in nearby source
free regions and, as shown by the errors on the ACIS-I net 
count rates (Table~\ref{infobs}), 
it contributes less than few percent to the total count rate.

\section{Chandra Imaging Analysis}

We have run a detection algorithm ({\it wavdetect} in CIAO) in
fields of view of $\sim$5$^{\prime}$$\times$2$^{\prime}$
of the ACIS-I3. Point-like emissions coincident with the optical position of the 
nucleus have been detected in all objects, confirming the X-ray/optical
association.  

We list the off-nuclear sources detected
within the optical extent (D25) of each galaxy
in Table~\ref{off}. We give the
positions in RA \& DEC (J2000), the distance from the nucleus in 
arcsec and kpc,
the number of net counts and relative errors,
the source significance and the 2-10 keV fluxes and luminosities
computed assuming a $\Gamma$=1.8 power-law spectrum modified by Galactic 
column densities, and the redshift of the corresponding galaxy.
We do not detect 
any off-nuclear source within the optical extent of IRAS\,01072+4954
(at the given exposure the flux limit corresponds to a luminosity
of 2.5$\times$10$^{39}$ erg/s).
All the off-nuclear sources are significantly detected at $>$ 3$\sigma$.
If they are actually associated with the host galaxies, 
the observed luminosities 
would be well in excess of the Eddington luminosity for a solar-mass black
hole or neutron star X-ray binary, therefore they should be considered
as Ultra-Luminous X-ray objects (ULXs,   Swartz et al. 2004).
Note that despite the presence
of these luminous sources, the galaxy X-ray emission is completely
dominated by the nucleus in each case.

Given the large RASS point spread function of about 45$^{\prime\prime}$
even for bright sources and the positional accuracy of
25$^{\prime\prime}$ (90\% confidence level, Voges et al. 1999), other
sources could in principle contaminate the RASS detection.
To verify whether any other X-ray source was present within the ROSAT
aperture we examined the $\sim 2^{\prime}$ region around each source.
We have found a source with $\leq 50$ cts at 121$^{\prime\prime}$ from the 
nucleus of IRAS\,01072+4954,
a source with $\leq 15$ cts at 55$^{\prime\prime}$ from
IRAS\,01319$-$1604 
and a source with  $\leq$ 45 cts   at
79.5$^{\prime\prime}$  from IRAS\,04392$-$0123.
In the present data, these sources contribute at most 15\% of the nuclear
flux, so that it is unlikely that they could have been dominant in 
the ROSAT observations.

\begin{figure}[h]
\begin{center}
\epsscale{.80}
\plotone{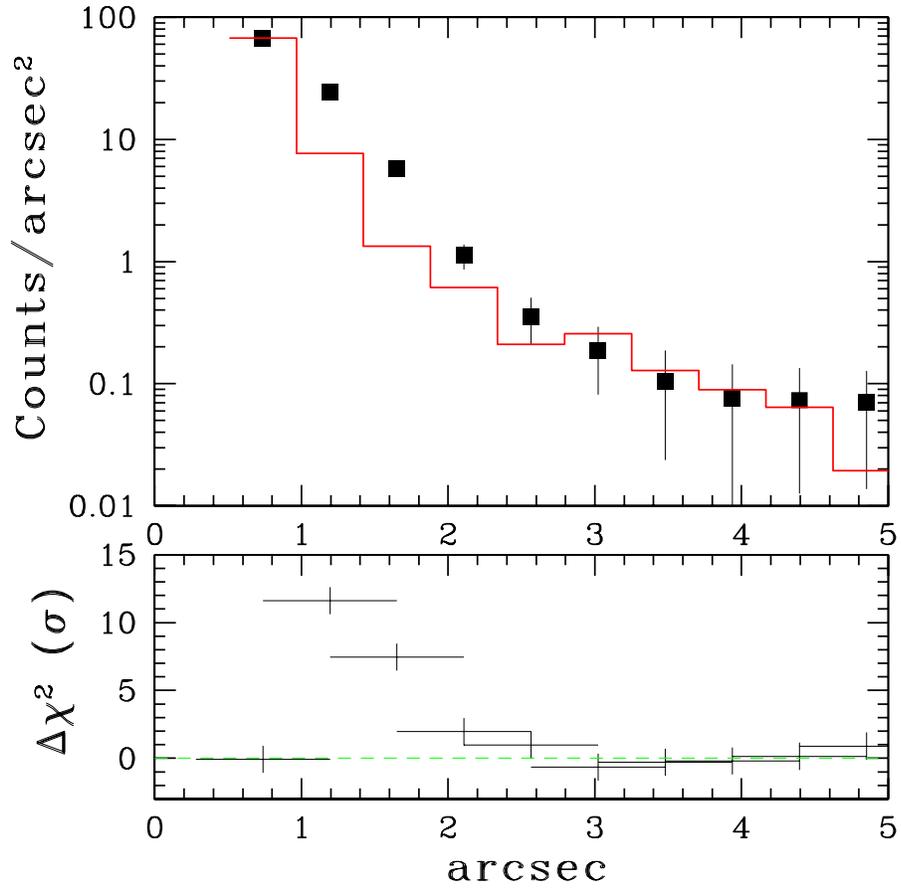}
\caption{IRAS\,01072+4954 radial profile (0.3-10 keV band) extracted from an annular region with 10 annuli 
within 5$^{\prime\prime}$. The solid line represents the {\it Chandra} PSF
simulated via {\em Chart}.}
\label{rprof}
\end{center}
\end{figure}

In order to look for extended emission, the radial profile of each
source has been derived in annular regions centered on the nuclear 
position
and reaching out to 5$^{\prime\prime}$, where the signal fades into
the background.  We also obtained a radial profile for the {\it Chandra}
PSF, by using the CIAO task {\em chart}, which simulates, by 
ray-tracing, a source
with the same spectrum as the target and an exposure time appropriate
to derive the statistics needed to define the PSF shape with virtually
no errors (we choose 100 ks).
After normalizing for the different exposure times, we compared
the source and the {\it Chandra} simulated PSF, as described in the CIAO
Threads, using the
SHERPA software. In all cases but one, the source is consistent with
a point source. 
In Figure~\ref{rprof} we show the radial profile of IRAS\,01072+4954; this
source is characterized by the presence of faint extended emission
between 1$^{\prime\prime} - 3^{\prime\prime}$ from the center, which correspond to a region of 0.5-1.3 kpc
(at the redshift of the source). 
This effect is more pronounced
in the soft (0.3-2 keV) than  in the hard (2-10 keV) energy band.

\section{Spectral Analysis}

Both {\it Chandra} and {\it Beppo}SAX spectral data 
have been grouped to at least 20 counts per
bin and spectra in the range of 0.3-10 keV have been analyzed using
the XSPEC software package. We first fitted {\it Chandra} and {\it Beppo}SAX spectra
separately. 
Each spectrum has
been initially fitted with a single power-law plus Galactic
absorption. In all objects this simple parametrization
describes well the observed spectra and the spectral shapes are within
the typical ranges observed for AGNs, i.e. photon indeces from
1.7 and 2, except for IRAS\,04392$-$0123 which shows a  flatter spectrum 
($\Gamma \sim 1.3$) both in {\it Chandra} and {\it Beppo}SAX observations. 
Intrinsic absorption values, when measured, are of the order of
10$^{21}$ cm$^{-2}$ or slightly in excess of the Galactic value
(Dickey \& Lockman, 1990). In the case of the flat spectrum of
IRAS\,04392$-$0123,
the upper limit derived on the column density (N$_{H}$ $<$
6$\times$10$^{20}$ cm$^{-2}$) remains consistent with the Galactic
absorption even when
fixing the spectral slope to the AGN canonical value ($\Gamma=1.9$).

\begin{figure*}[htb]
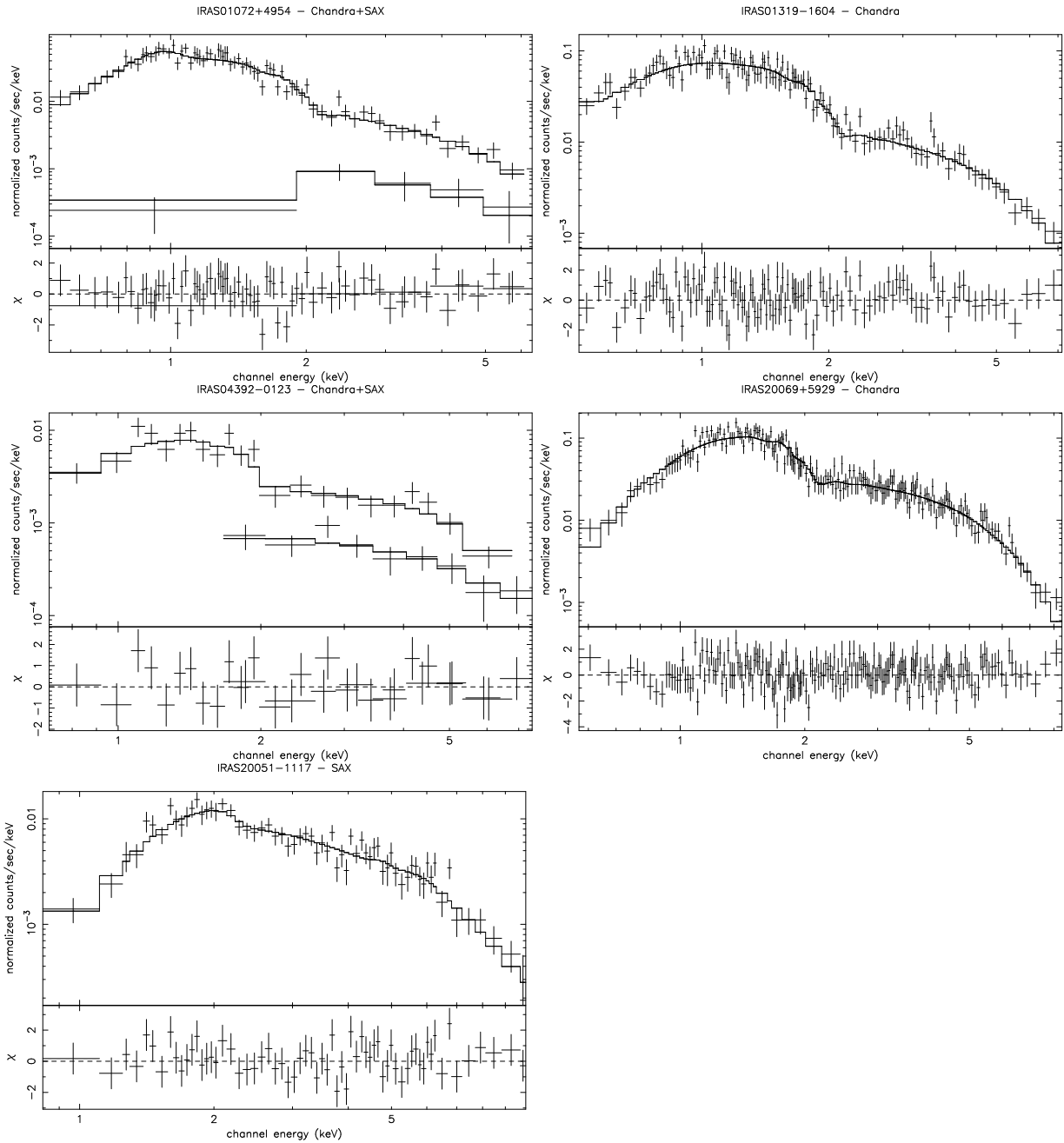

\begin{center}
\vbox{
\hbox{
\psfig{file=f2a.eps,width=8truecm,angle=-90}
\hfill
\psfig{file=f2b.eps,width=8truecm,angle=-90}
}
\hbox{
\psfig{file=f2c.eps,width=8truecm,angle=-90}
\hfill
\psfig{file=f2d.eps,width=8truecm,angle=-90}
}
\psfig{file=f2e.eps,width=8truecm,angle=-90}
}
\caption{Top left: IRAS\,01072+4954 - {\it Chandra} and {\it Beppo}SAX spectra
fitted simultaneously with a power-law, and free relative calibration.
Top right: IRAS\,01319$-$1604 - {\it Chandra} spectrum modeled with a soft 
thermal component plus an unabsorbed power-law.
Middle left: IRAS\,04392$-$0123 - {\it Chandra} and {\it Beppo}SAX spectra fitted 
simultaneously with a power-law, and free relative calibration.
Middle right: IRAS\,20069+5929 - {\it Chandra} spectrum modeled with an
absorbed power-law. Bottom: IRAS\,20051$-$1117 - {\it Beppo}SAX spectrum modeled with a 
power-law and low energy absorption.
}
   \label{spectra}
\end{center}
\end{figure*}

{\it Beppo}SAX spectral results are listed
in Table~\ref{bep}. 
Although at lower significance, 
the {\it Beppo}SAX spectral parameters are in agreement within the
errors with those derived from the {\it Chandra} data. 
We have therefore fitted all the available data for the same source together.
The final spectra are plotted in Figure~\ref{spectra} while the
spectral results are presented in Table~\ref{spec}. We also report the
results from Georgantopoulos, Zezas \& Ward (2003), Georgantopoulos et
al. (2004) in the same table for ease of reference.
The quoted
errors on the spectral parameters correspond to the 90\%
confidence level for one interesting parameter.

We discuss here each source in turn.
IRAS\,20051$-$1117 is the brightest source of the sample observed by {\it
Beppo}SAX. Its spectrum is well described by a simple power-law, with a
photon index ($\Gamma$$\sim$1.9) in agreement with the one
reported by Georgantopoulos et al. (2004), using {\it Chandra} and
XMM-{\it Newton} data. The addition of a Gaussian component consistent with
an FeK$\alpha$ line (at 6.29$^{+0.28}_{-0.48}$ keV, rest frame) is 
significant at only 2$\sigma$ (EW = 282$^{+215}_{-214}$ eV), 
again in agreement with  Georgantopoulos et al. (2004)
{\it Chandra} results. Residuals are also present around
6.9 keV (see Figure \ref{spectra}) but at a lower statistical significance.
At high energies (13-200 keV), the PDS data 
fall above the simple power-law extrapolation from the MECS (1.3-10 keV), 
by a factor of $\sim 5$.
We explored the $\sim 1^{\circ}$ region around the target
to look for possible contaminant sources. 
We found 1RXS\,J200433.5-111345, detected by RASS at $\sim 50^{\prime}$
from IRAS\,20051$-$1117 and brighter by a factor 1.3,
corresponding to a galaxy detected by IRAS (no other relevant information
like redshift, morphological classification, etc. is available). 
The source could be the
contaminant if it hosts an absorbed nucleus. Given the discrepancy in the
relative fluxes, the partial contamination from 1RXS\,J200433.5-111345
or other sources is likely, and therefore 
we make no further use of the PDS spectrum.
On the other hand a Compton thick AGN in IRAS\,20051$-$1117
is ruled out by the lack of a FeK line with high EW (see below).

IRAS\,01072+4954 and IRAS\,04392$-$0123 have been observed by both 
{\it Chandra} and {\it Beppo}SAX. 
In Figure~\ref{spectra} we show {\it Chandra} and {\it Beppo}SAX spectra
fitted simultaneously leaving the relative normalization
free to vary. Their values are in the range 1-1.3, indicating that
the {\it Chandra} and {\it Beppo}SAX normalizations
are within 30\%, that would either indicate a minimal flux variation
between the two observations, or the uncertainty in the relative calibration
between instruments.

In the {\it Chandra} spectrum of IRAS\,01072+4954 a thermal component (MEKAL) has
been added to the model to account for residuals visible below 2 keV
above the simple power law.
Such component has a 0.3-2 keV flux of 4.6$\times$10$^{-14}$ erg 
cm$^{-2}$ s$^{-1}$
(L$_{0.3-2 keV}$ $\sim$ 5.8$\times$10$^{40}$ erg s$^{-1}$) and it is
significant at 99.99\% (via an F-test),
consistently with the
presence of the faint extended soft component revealed by its radial
profile.

IRAS\,20069+5929 has the highest fitted value for intrinsic N$_{\rm H}$,
which is however still consistent with absorption from the host galaxy.

The presence of an Fe line at 6.4 keV is not statistically
significant in any of the sources observed by {\it Chandra}.
In the case of IRAS\,00317$-$2142 and IRAS\,01072+4954 the upper limit
to the equivalent widths at 90\% (as shown in Table~\ref{spec})
is not stringent, due to the poor statistics above 6 keV.
In the other cases, 
the upper limits 
are always consistent with expectation from the low intrinsic absorption 
measured, ruling out the
Compton thick hypothesis (Bassani et al. 1999).

\section{Timing Analysis}

\subsection{Short-term variability}

\begin{figure}[h]
\vbox{
\hbox{
\psfig{file=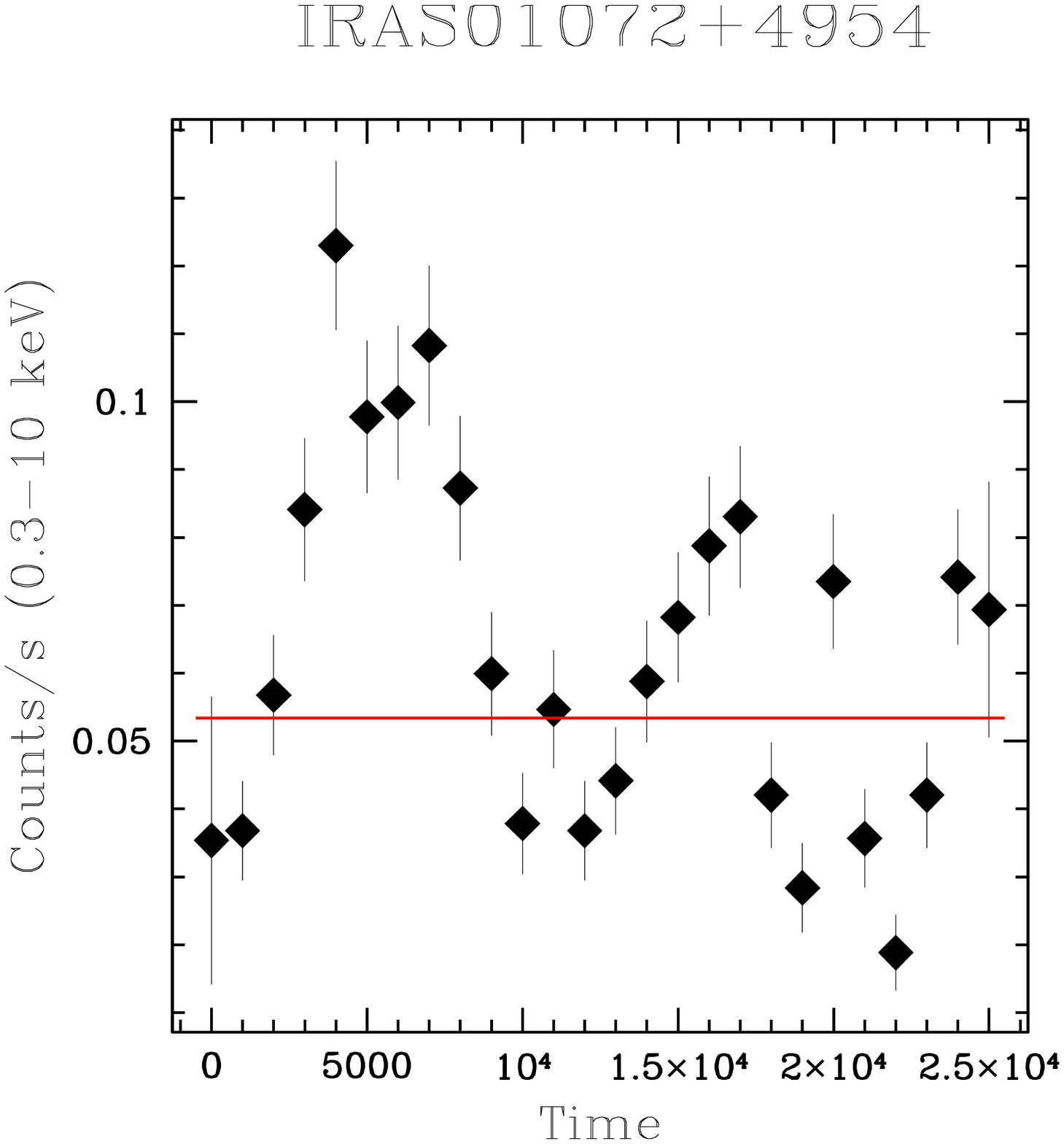,width=4truecm}\hfill
\psfig{file=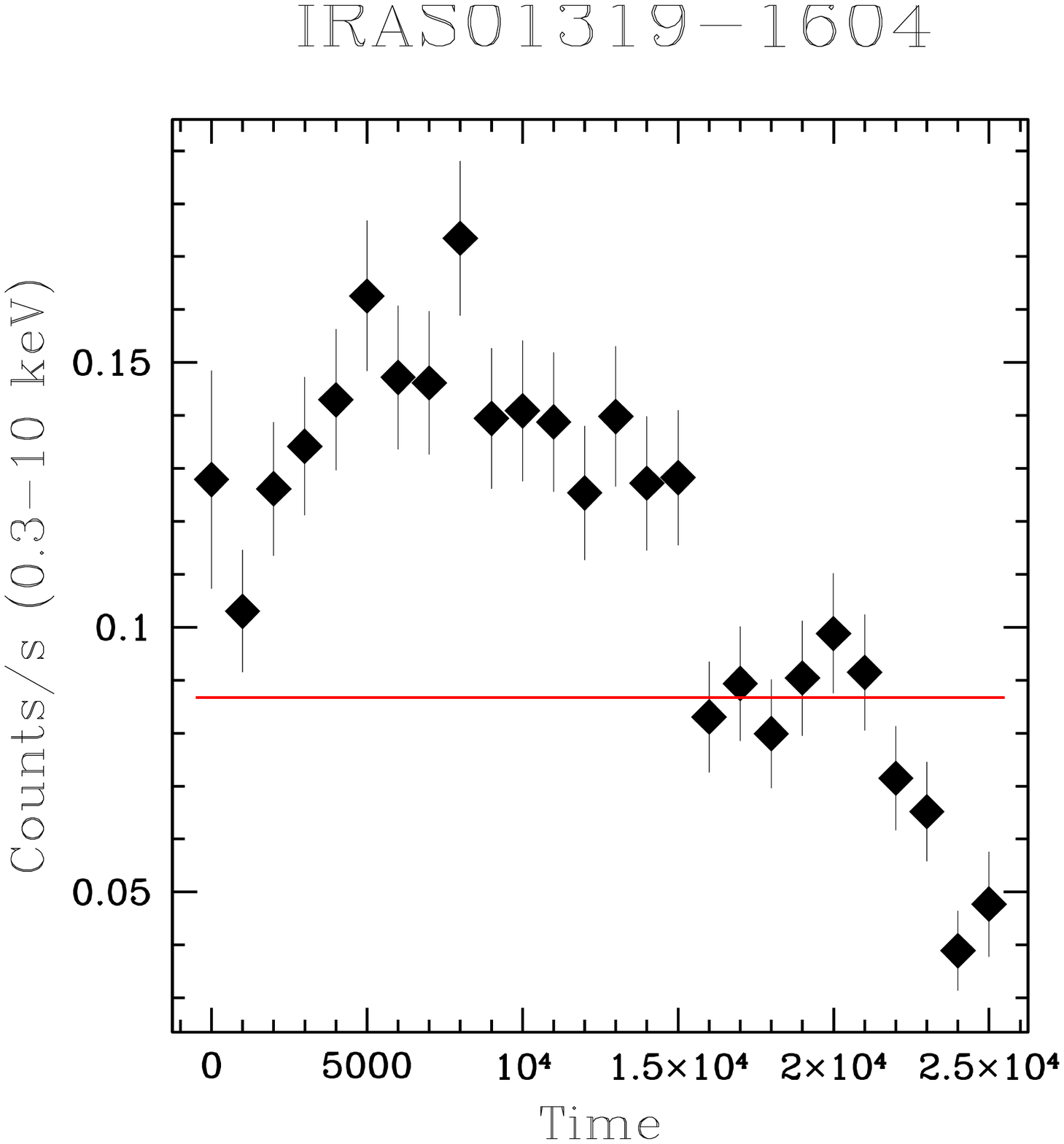,width=4truecm}
}\vfill
\hbox{
\psfig{file=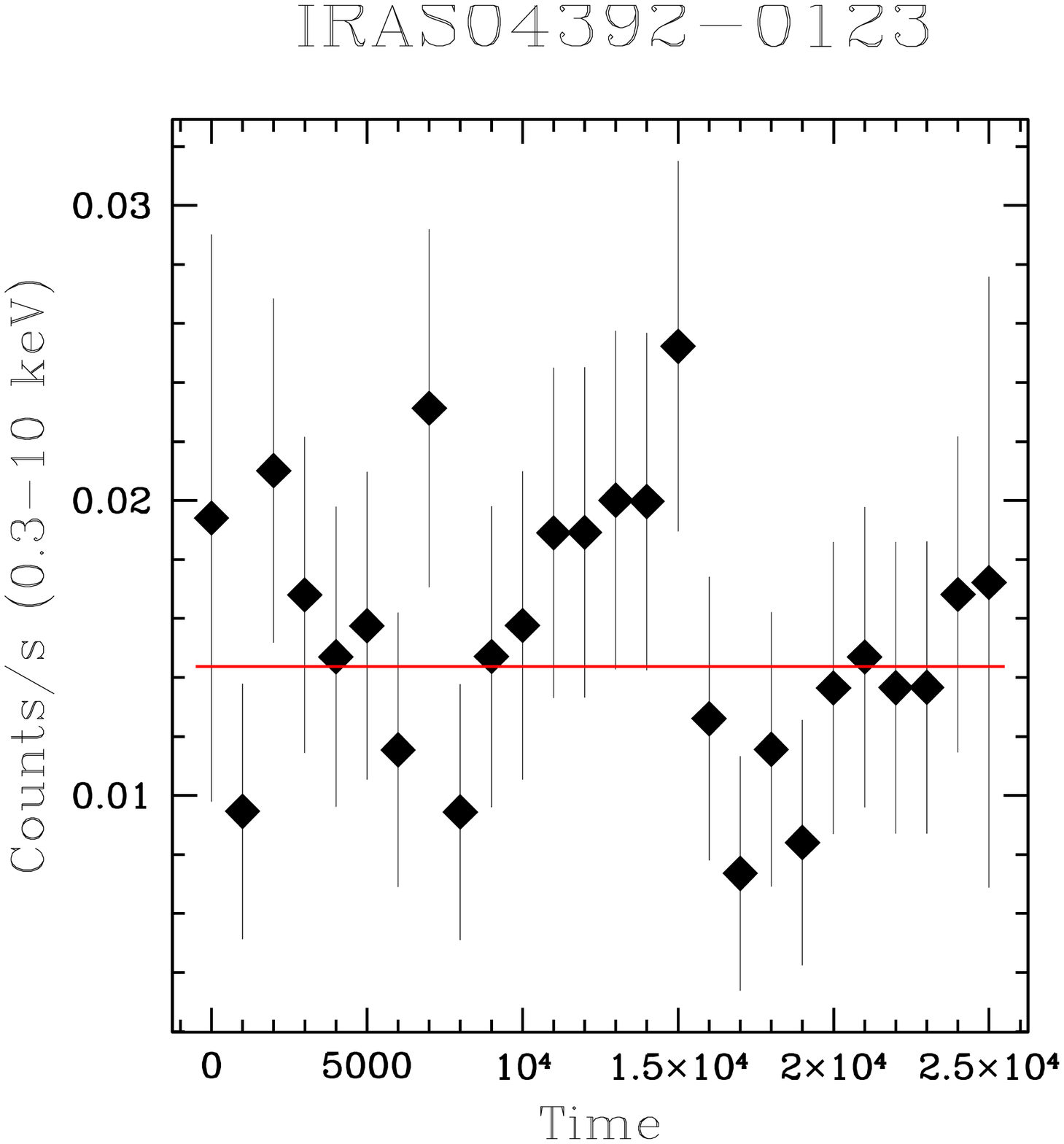,width=4truecm}
\hfill
\psfig{file=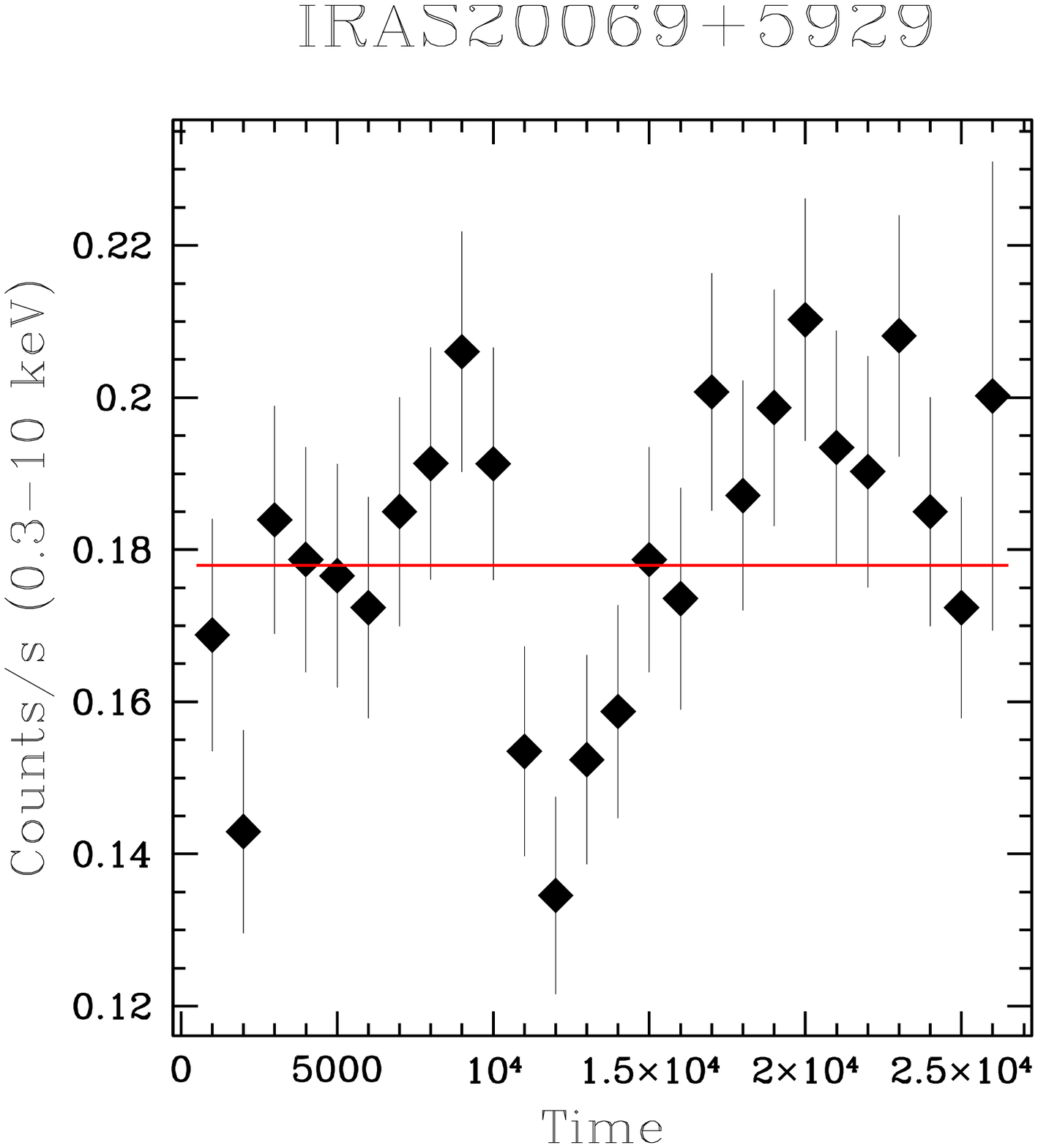,width=4truecm}
}
}
\caption{{\it Chandra} light curves in the 0.3-10 keV energy band binned at 1 ks
(Top left: IRAS\,01072+4954, Top right: IRAS\,01319$-$1604;
Bottom left: IRAS\,04392$-$0123; Bottom right: IRAS\,20069+5929).}
\label{varcha}
\end{figure}

\begin{figure}
\hbox{
\hfill
\psfig{file=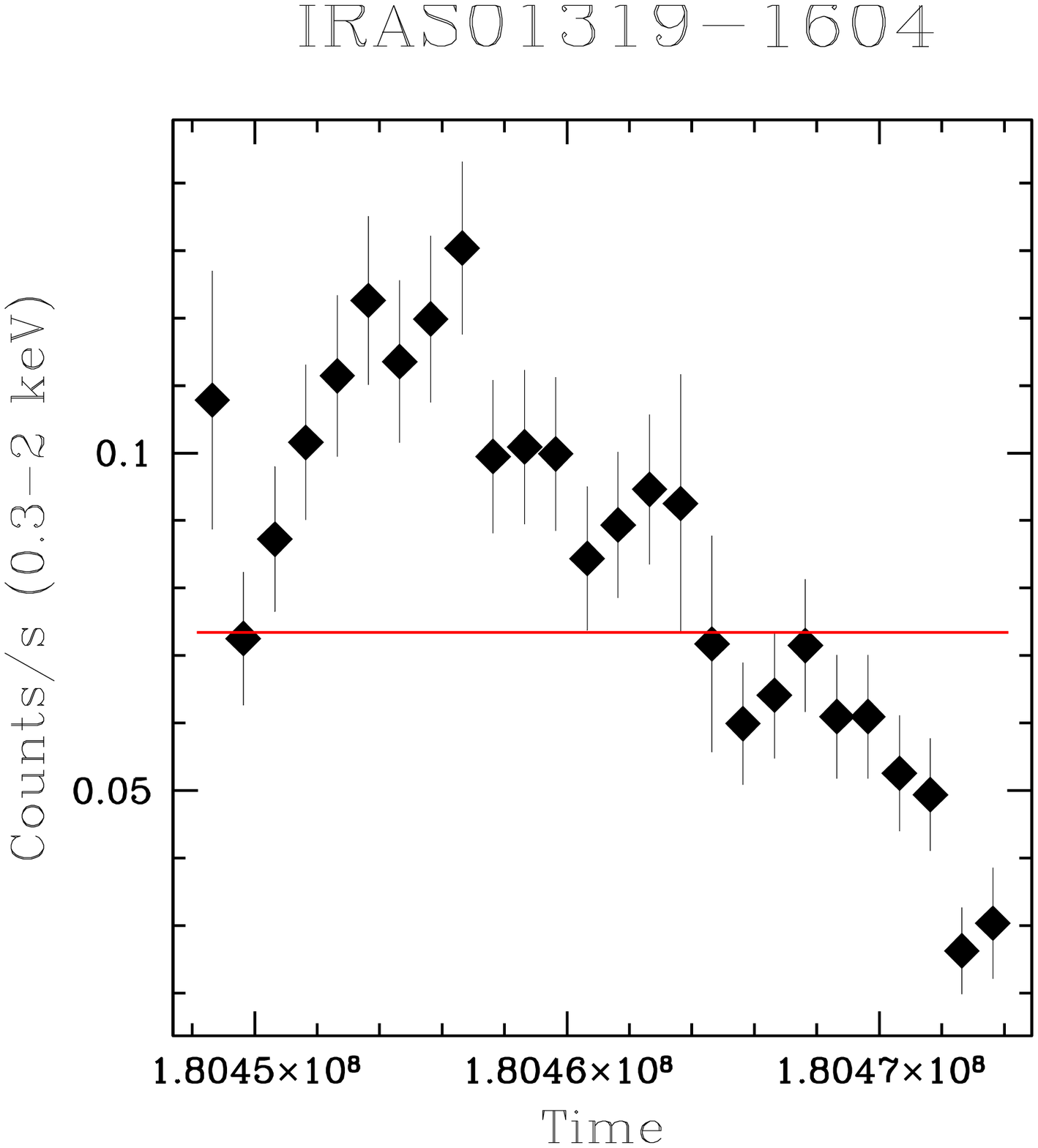,width=4truecm} \hfill
\psfig{file=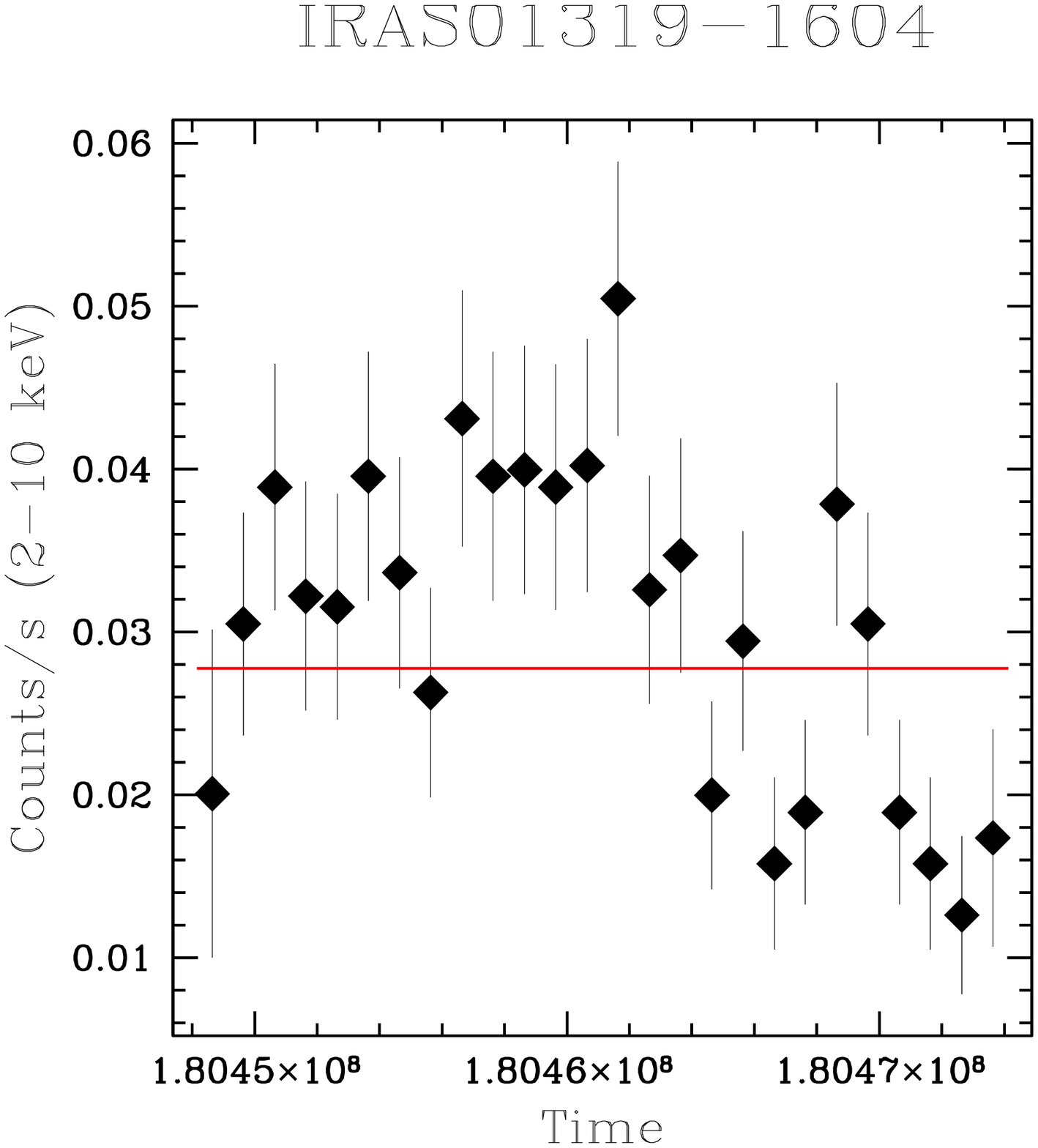,width=4truecm}
}
\caption{{\it Chandra} light curve in the soft and hard band for IRAS\,01319$-$1604.}
\label{lc:soft+hard}
\end{figure}

\begin{figure}
\begin{center}
\includegraphics[angle=0,scale=.2]{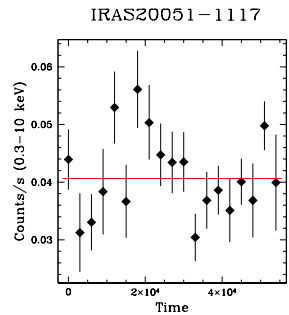}
\caption{{\it Beppo}SAX light curve for IRAS\,20051$-$1117 in the 0.3-10 keV 
energy band binned at 3 ks.}
   \label{varsax}
\end{center}
\end{figure}

The ACIS-I light curves have been examined in order to look for short
term variability. In Figure~\ref{varcha} we show the 0.3-10 keV
Chandra light curves of all sources. They have been extracted from circular
regions of 3$^{\prime\prime}$ radius,
binned at 1 ks and fitted with a
constant. The resulting constant values and $\chi$$^{2}/dof$ are given
in Table~\ref{vartab} for the 0.3-10 keV, 0.3-2 keV and 2-10 keV energy
ranges. In the cases of IRAS\,01072+4954, IRAS\,01319$-$1604 and
IRAS\,20069+5929 the hypothesis of a
constant flux is rejected at the 99\% confidence level. In
particular, IRAS\,01072+4954 and IRAS\,01319$-$1604
show larger flux
variations in the soft X-ray band.
To illustrate this point, in Figure~\ref{lc:soft+hard}
we plot the soft and hard light curves
for IRAS\,01319$-$1604, the most variable source of the sample.
The non-detection of
variability in IRAS\,04392$-$0123 is possibly due to the low statistics 
available.
We examine also the {\it Beppo}SAX light curve of IRAS\,20051$-$1117 (see
Fig.~\ref{varsax}), in bin sizes
of 3 ks. Fitting a constant to the observed count rate we
reject the constant flux hypothesis at more than 95\% level
(see Table~\ref{vartab}). 
However, {\it Chandra} or XMM-{\it Newton} short-term light curves
show no variability either in this source (Georgantopoulos et al. 2004),
nor is detected in {\it Chandra} for 
IRAS\,00317$-$2142 (Georgantopoulos, Zezas \& Ward 2003).
The variability of a factor 2-3 for IRAS20051-1117 over a time scale of $\sim $4 ks implies 
(using light crossing arguments) that the dimensions of the
emitting region are typical of a nuclear source
(${\leq} 1.2\times 10^{14}$ cm, i.e. $4\times 10^{-5}$ pc).

When the statistics allows it, we have attempted to extract the spectrum 
at different intervals to check for possible spectral variations, suggested
by the higher variability measured in the soft band.
However, 
the spectral parameters measured in the different states (high and low flux)
are always consistent within errors.

\subsection{Long-term variability}

\begin{figure}
\psfig{file=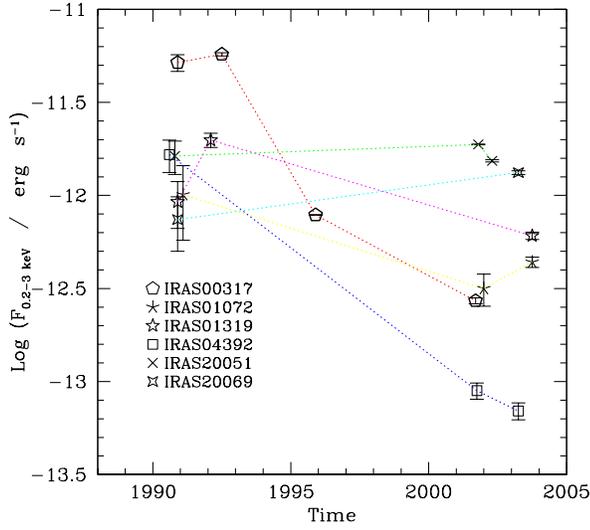,width=8truecm}
\caption{
Long term trends for the 6 sources in the sample in the
0.3-2 keV band. All the fluxes reported are unabsorbed and recomputed
assuming the {\it Chandra} spectral best-fit. Points in the
1990-1992 years are from ROSAT, in 1995 from ASCA 
and 2001-2003 from {\it Beppo}SAX and {\it Chandra}.}
\label{long_fig}
\end{figure}

In order to examine the presence of long-term variability
we compared the observed fluxes from all available X-ray measurements, both
from literature and from our {\it Beppo}SAX and {\it Chandra} data presented here.
 Figure~\ref{long_fig} shows the long term light curves for the six sources
in the sample. Errorbars have been plotted for all the fluxes
except those derived from the literature for which errors on the count rates
were not available.
For a better comparison, all ROSAT, {\it Beppo}SAX and {\it Chandra} 0.3-2 keV fluxes
have been recomputed,
assuming the {\it Chandra} best fit spectral shape
as reported in Table~\ref{spec}. The resulting
0.3-2 keV unabsorbed fluxes are given in Table~\ref{long_tab}.
All our sources have been observed by the ROSAT All-Sky Survey (RASS)
with the Position Sensitive Proportional
Counter (PSPC) in the 0.1-2.4 energy band. The RASS count rates
and errors have been taken from Boller et al. (1992). 
IRAS\,00317$-$2142 and IRAS\,01319$-$1604 have also been observed in a
PSPC pointed observation.
In the 2 yr period between the RASS and the PSPC pointed observation, 
IRAS\,00317$-$2142 does not show significant variability 
while IRAS\,01319$-$1604 has nearly doubled its flux (measured by the same 
instrument). ASCA and {\it Chandra} count rates of IRAS\,00317$-$2142 have been taken from
Georgantopoulos (2000) and Georgantopoulos, Zezas \& Ward (2003), 
respectively. In the case of IRAS\,20051$-$1117,
Chandra and XMM-{\it Newton} count rates, taken from
Georgantopoulos et al. (2004), were both obtained
on 2002 April 1 and the measured soft fluxes are comparable
within a few percent; therefore we only consider the {\it Chandra} data for the long-term analysis.

Unfortunately, the statistics in the ROSAT observations 
is not sufficient to derive a reliable spectral measurement, therefore
a possible spectral variation could have occurred
between the ROSAT and the {\it Beppo}SAX/Chandra epochs, even if
we do not observe spectral variability
between the {\it Beppo}SAX and {\it Chandra} observations.
While we do not observe flux variations on time scales of $\sim$ 2 yr 
between the
{\it Beppo}SAX and the {\it Chandra} observations, IRAS\,04392$-$0123 has experienced a
variation by a factor of $\sim$ 24 in $\sim$ 10 yr between the ROSAT
and the {\it Chandra} observations, similarly to the case of IRAS\,00317$-$2142
which varied by a factor of $\sim$ 20 (see also Georgantopoulos, Zezas \& Ward, 2003). Also
IRAS\,01072+4954 and IRAS\,01319$-$1604 varied, even though only by a
factor of $\sim$ 3.
The strong long term variability in the X-rays seems 
very common in this class.

\section{Discussion}

The X-ray analysis of Composite galaxies has
revealed their AGN dominance in this spectral domain. 

In what follows we make use of various diagnostic diagrams
and multi-frequency data to investigate the reasons for the
contradictory optical/X-ray classification of these objects, 
and suggest a global explanation of the phenomenon.

\subsection{Diagnostic Diagrams}

\begin{figure}[ht]
\vbox{
\includegraphics[scale=0.45]{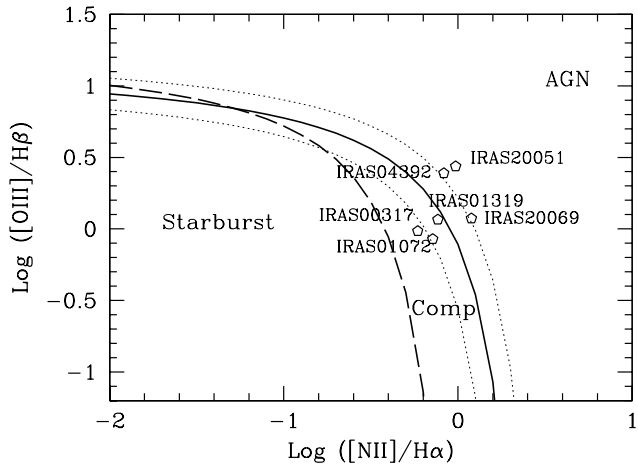}
}
\caption{Veilleux and Osterbrock (1987) diagnostic diagrams.
The optical emission line ratios are from Moran
et al. (1996) and have been corrected from Galactic extinction. 
 The solid line is the theoretical dividing line from Kewley et al. (2001)
(together with the dotted lines which indicate its error range);
an updated dividing (dashed) line from Kauffmann et al. (2003)
has also been plotted. 'Comp' indicates
the region of the diagram where Composite objects are
expected to be found.}
\label{vo}
\end{figure}

The optical diagnostic diagrams obtained using 
standard emission line ratios (Veilleux \& Osterbrock, 1987)
show that Composites are very close to the boundary region between
starbursts and AGNs confirming the M96 classification. 
This is shown in Figure~\ref{vo} where 
the [O~III]/H$\beta$ ratio is plotted 
vs. the [NII]/H$\alpha$ (the other two standard
diagnostics which make use of [SII]/H$\alpha$ and 
[OI]/H$\alpha$ ratios give similar results).
The lines represent the theoretical
starburst limits, a standard one which have been taken from 
Kewley et al. (2001) (together with the dotted lines which indicate the error range)
and an updated estimate for the
starburst boundary derived from the SDSS observations (from Kauffmann
et al. 2003).
The location of the `Composite' region should be between these two lines
(Hornschemeier et al. 2005).

\begin{figure}[h]
\begin{center}
\psfig{file=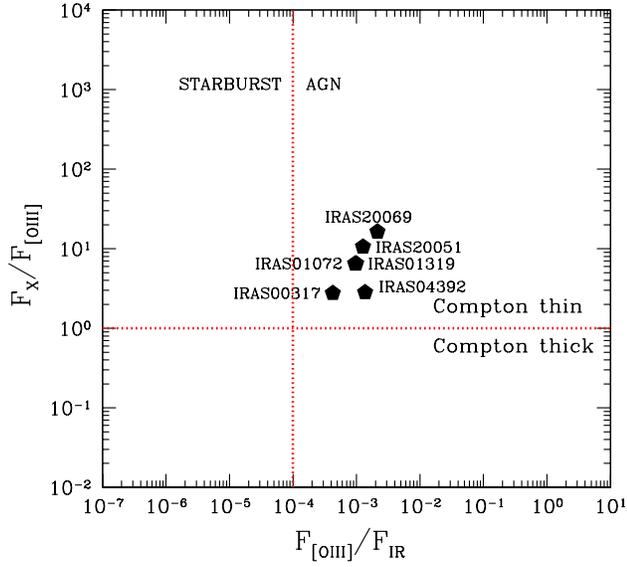,width=8.5truecm}
\caption{F$_{X}$/F$_{[O~III]}$ vs. F$_{[O~III]}$/F$_{IR}$.  The
fluxes are listed in Table~\ref{sample}. The [O~III]$\lambda$5007 flux
of each galaxy has been corrected for the relative extinction using
the formula given in Bassani et al. (1999).
Compton thin, Compton thick and starburst regions (as derived in Panessa 
\& Bassani 2002) are separated by dashed lines.}
\label{dia}
\end{center}
\end{figure}

However, when the flux of the optical emission line [O~III]$\lambda$5007
is combined with the infrared and X-ray fluxes, Composite galaxies
are classified as AGNs rather than starburst.
In Figure~\ref{dia} the diagram with the combination of 
F$_{X}$/F$_{[O~III]}$ vs. F$_{[O~III]}$/F$_{IR}$ ratios
is shown. These flux ratios have been used to separate 
the AGN and the starbursts contributions by
Panessa \& Bassani (2002) and Braito et al. (2004), based on the fact that 
the [O~III]$\lambda$5007 flux is associated with the AGN and
the far-infrared emission is associated mainly with the star-forming
activity. At the same time they are a powerful tool in the detection
of heavy obscuration not seen in X-rays below 10 keV (as in Compton
thick objects, Bassani et al. 1999). 
Clearly Composite objects all fall in the AGN region, in good
agreement with the relative 
broadness of the [O~III]$\lambda$5007 line found by M96
which points to an AGN origin.
The diagram further shows that they should all be classified as Compton thin
AGN (note that the X-ray fluxes used in the plot are those observed
in the `low-state' epoch, and therefore the classification 
would be also valid in the 'high-state').
This indicates very little amount 
of absorption, in agreement with the results obtained from the 
X-ray analysis, i.e. the absence of a strong Fe-line and X-ray obscuration.
Therefore the relative weakness of the AGN optical
emission lines cannot be explained as due to the presence of absorbing
material in the line of sight, as  suggested by M96.

\subsection{X-ray vs. infrared luminosity diagram}

\begin{figure}
\begin{center}
\psfig{file=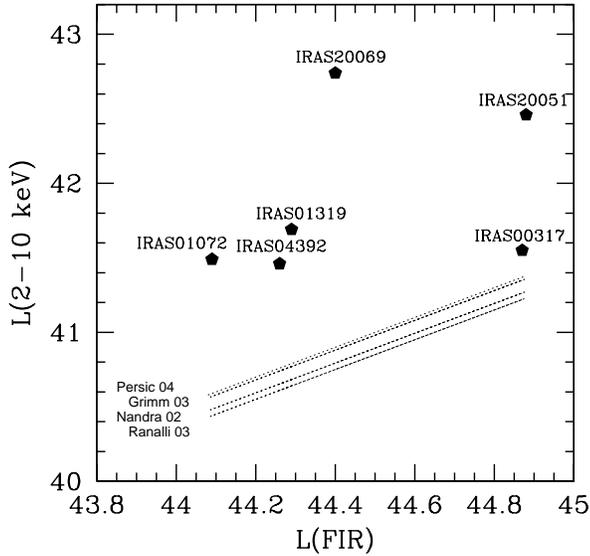,width=8truecm}
\caption{Chandra 2-10 keV 
luminosity vs. far-infrared luminosity
for the Composites. Lines represent expected X-ray luminosity from the
starbursts, following different authors, as explained in the text.
}
\label{sfr}
\end{center}
\end{figure}

Having assessed the importance of the AGN component from the X-ray,
infrared and [O~III] emissions combined, 
we want to estimate here the amount of X-ray emission expected to be
produced by the presence of a starburst. Therefore we derived
the Star Formation Rate (SFR) from the far-infrared luminosity (Kennicutt 1998),
assuming that the latter (given in 
Table~\ref{sample}) is all due to the starburst.
The SFRs obtained are in the range of 5-34 M$_{\odot}$/yr
(note that these values are upper limits 
since we have not subtracted the possible AGN contribution from the FIR luminosity).
Subsequently, the L$_X$ in the 2-10 keV band associated to the 
derived SFRs have been estimated as in Nandra et al. (2002), 
Grimm et al. (2003), Ranalli et al. (2003), Persic et al. (2004).
%and also the L$_X$ in the 0.5-2 keV band from Ranalli et al. (2003).
The derived correlations have been plotted in  
Figure~\ref{sfr}, together with the 
observed hard 
%(left panel) and soft (right panel) 
L$_X$ vs. L$_{FIR}$ for each Composite object.
For all sources, regardless of the correlation considered, 
the X-ray observed emission is well above the
value expected to be produced by a starburst.
%both in the hard and in the soft X-ray bands.} 
The SFR derived here for the Composites correspond
to a starburst that is not bright enough to
produce the observed X-ray emission, which is then mostly given by the
AGN, as suggested by our X-ray analysis.
Observed X-ray emission in excess to that predicted from SFR 
has also been observed in high-redshift
submillimeter sources (Alexander et al. 2003),
that might point to the presence of
an active nucleus also for this class of objects.
Only in the case of IRAS\,00317$-$2142 the observed X-ray luminosity is
close to the expected value for a starburst; this is in agreement with the results
of the multi-wavelength analysis for this object presented in the next
Section.

\subsection{Spectral Energy Distributions}

Spectral Energy Distributions (SEDs) have been assembled from radio
($\nu\approx10^8$ Hz) to hard X-rays ($\nu\approx10^{18}$ Hz) for our
sample sources. Radio, far to near infrared, optical, X-ray data have
been taken from NED, and complemented with the {\it Chandra} data from this
work.  All X-ray data points have been plotted with the spectral
shape measured in the {\it Chandra} observations.

In Figure~\ref{sed1} we compare the observed SEDs with the templates of
Medium Energy Distribution for radio quiet quasars (Elvis et al. 1994),
starburst galaxies (Schmitt et al. 1997) and normal spiral galaxies
(Elvis et al. 1994). 
The templates we show are normalized to match, and not exceed data points. 
In particular, the starburst template is
normalized to the radio-IR portion of the spectrum, while the X-ray one
is normalized to the different X-ray states observed.  
We stress that we are only interested in
deriving an overall consistent picture, without attempting
a quantitative computation of the contributions of the
different components, that would require at least simultaneous
data to account for the observed variability in the X-ray band. 
Moreover, it must be taken into account
that all data points have been taken using different
apertures, in particular for the optical band where 
the emission is heavily contaminated by the host galaxy. 
The SEDs show clearly the {\it Composite} nature of these
objects: the AGN dominates at X-ray wavelengths, while
the starburst is the most
important contributor to the mid$/$far IR emission; the host
galaxy template accounts for the optical appearance. 

It is evident that, in the optical band, the
AGN contribution is always less than that of the starburst, and, 
except for a couple of sources during their ``high'' state,
even by a factor of 10. 
Note instead that in IRAS\,00317$-$2142, when it is in the low state, 
the contributions to the X-ray 
emission from the AGN and the starburst become comparable (in agreement with
Figure~\ref{sfr}).

The AGN contributes to the total bolometric luminosity from
less than 10\% in the 'low flux state', to 15-30\% in the 'high flux state'
thus making the SB contribution dominant in the bolometric output, except
possibly during bright AGN flares.

Mid-infrared and L-band spectroscopy could provide a powerful way to
disentangle the starburst from the AGN component (Genzel et al. 1998,
Imanishi 2002, Risaliti et al. 2003, Lutz et al. 2003).
Moreover broad band data from the near to the far-infrared
frequencies could provide a more detailed characterization
of the SEDs, in particular by exploiting the 
Spitzer Space Telescope unprecedented capabilities 
(see e.g. Franceschini et al. 2005).

\subsection{A weak and low M$_{BH}$ AGN?}

IRAS\,00317$-$2142, IRAS\,20051$-$1117 and IRAS\,20069$+$5929 are 
the only three sources for which a weak broad component of the
H${\alpha}$ emission line has been detected in their optical spectra,
while in all objects the narrow H${\alpha}$ component is probably due
to the starburst. 
A correlation between the 2-10 keV X-ray luminosity versus 
the H${\alpha}$ luminosity has been widely observed in high and low
luminosity AGNs (Ward et al. 1988, Ho et al. 2001).
IRAS\,00317$-$2142, IRAS\,20051$-$1117 and IRAS\,20069$+$5929,
for which we consider only the broad H${\alpha}$ components, follow
the Ho et al. (2001) correlation 
(which applies to both high and low luminosity AGNs).
The remaining three Composites for which the
broad H$\alpha$ was not measured are generally fainter 
both in X-rays and in the optical band; 
therefore it is likely that an optical emission line
luminosity of a factor of 2-3 lower could be difficult to measure. 
 Using the Kennicutt (1998) relation, we derived the SFR from the H${\alpha}$ luminosity
produced by the narrow component and we estimated the expected 2-10 keV luminosity to be
produced by that SFR (Ranalli et al. 2003). The expected 2-10 keV luminosities are
in the range of 10$^{39-40}$ erg s$^{-1}$ , i.e. a factor of 50-100
lower than the observed ones. 
This result is consistent with our previous interpretation that
in all objects the narrow H${\alpha}$ component is probably due
to the starburst, and the X-rays to the AGN.

For the three objects for which we have a measure of the broad H$\alpha$
component, we attempt an estimate of the Black Hole Mass.
We make use of the
formula $M_{BH} = v^2 R_{BLR} / G$ (McLure \& Dunlop 2001) and of a few
additional assumptions.
The quantities $ v = 1.5 \, FWHM \, H_{\beta}$
and $R_{BLR} = 32.9 (\lambda L_{5100} / 10^{44}$ erg s$^{-1}$)$^{0.7}$
in light days (Kaspi et al. 2000) require a measure of the broad $H_{\beta}$
emission, which is not detected because swamped
by the narrow line produced by the starburst
and of the AGN continuum, which we do
not measure directly.
We substitute the FWHM of H$\beta$
with the FWHM of the broad H$\alpha$ that we
attribute to the AGN (from M96).
If anything, the H$\alpha$ is usually broader than
H$\beta$, therefore we will overestimate the resulting mass.
We infer the AGN continuum from the
template fitted to the X-ray luminosity. This is a rough approximation of the
real illuminating continuum, but it should be correct within an order
of magnitude at least.
Plugging these numbers in the above formula we derive numbers for the mass
of the order of a few 10$^{5-6}$ M$\odot$. An order of magnitude uncertainty
on the L$_{5100\AA}$ changes the mass by 10$^{0.7}$. We therefore can conclude
that most probably the masses of these black holes are small
and that they could possibly undergo strong changes in accretion rate when they brighten.
We have found a possible analogy between our sources and the
Seyfert 1.5 NGC 4395.
It is a low luminosity AGN with a small
black hole mass, that shows X-ray variability, although
more violent than what observed in the Composites (Vaughan et al. 2005).

The Black Hole Mass range is consistent with 
all these objects being late type galaxies
(as indicated by their morphological types in Table~\ref{sample}) with small
stellar bulges.
In fact, from their total B-band magnitude and an average bulge/total flux
ratio typical of their morphological type (de Jong 1996), 
we derived the black hole mass values
which are consistent with those previously found, 
when assuming that the $L_{B,bulge}-M_{BH}$ correlation
(e.g., Yu \& Tremaine 2002) still holds for late type spirals.

Detailed follow up observations (e.g.
high spatial resolution optical spectroscopy) are needed in order to provide
a more reliable estimate of the black hole mass
for these objects.

\section{Conclusions}

New {\it Chandra} and {\it Beppo}SAX observations 
of 4 and 3 Composites, respectively, have deepened our knowledge
of this class of sources.
Based on the X-ray analysis presented here, Composite galaxies behave
like typical type 1 AGNs in the X-ray band: their emission is dominated by a
bright nuclear source, whose spectral properties are
typical of this class. A power-law model with spectral
index ($\Gamma$ = 1.7-2.1) and little intrinsic absorption (N$_{H}$
$<$ 4$\times$ 10$^{21}$ cm$^{-2}$) well describe their spectra. 
Iron lines are not significantly
detected (upper limits on their equivalent widths are below $\sim$ 400
eV). Large flux variability is found on many different temporal scales. 

A broad analysis of the whole Composite sample, carried out by adding to
our X-ray data the results from Georgantopoulos (2000), 
Georgantopoulos, Zezas \& Ward (2003), Georgantopoulos et al.
(2004) and multi-wavelength data from the literature, has
revealed that the study of this class is relevant both 
for the investigation of the AGN-starburst connection 
and for the X-ray properties in medium/low luminosity
AGNs.

Interestingly enough, AGN and starburst
activity seem to be present with almost the same intensity in this class of objects.
Spectral Energy Distributions have clearly shown that
the different emission components contribute to different spectral
energy bands: the infrared emission is 
probably dominated by the starburst and the X-ray one
derives from the AGN. The optical continuum 
is mainly contaminated by the host galaxy light
while the emission line spectrum shows narrow emission lines
produced by starburst. The H${\alpha}$
emission line is probably associated to both: the narrow component
to the starburst and the broad component, 
when bright enough to be detected, 
correlates well with the hard X-ray luminosity, 
and therefore is to be ascribed to photoionization by the AGN.

It is unlikely that heavy obscuration
that could explain the weakness of
the AGN in the optical band is present, as shown by 
the pronounced X-ray variability 
and low column densities and by the flux diagnostic diagrams
shown in section 7.1. 
A dusty clumpy ionized absorber, able to selectively
obscure the optical emission, that leaves the X-ray emission almost
unabsorbed (Georgantopoulos 2000; Maiolino et al. 2001) is clearly possible.
However the detection of broad H${\alpha}$ with FWHM typical
of type 1 objects makes this hypothesis remote.  In conclusion,
the lack of clear indications of the presence of an AGN in the
optical spectra of the Composites is probably due to a combination
of the faintness of the AGN itself and of the masking effect of
the starburst.

As mentioned above, the most striking characteristic of Composite galaxies
is the strong, long- and short-term, X-ray variability observed.
ROSAT observation taken between 1991 and 1992
reveal bright soft nuclei at the level of 10$^{42-43}$ erg/s.
{\it Chandra} and {\it Beppo}SAX observations taken nearly 10 years
later reveal that, in most cases, these objects are fainter. 
Two objects show a variation
in flux by a factor of 20, while in two other objects  the fluxes 
decreased by a factor of 2-3.
It is likely that the inclusion of these sources in the
RASS-IRAS correlation of M96 is partly due to their
bright state at that time.
What is the cause for their brightening?
Are they now fading away or just changing state every once in a while?

Concerning the flux variation of a factor of $\sim$20 over $\sim$10
yrs, we cannot say much: it is not an extreme value, but certainly 
not the most common (see eg. Ulrich et al 1997).  Furthermore, the data
sampling is very scanty and does not allow us to determine the
time dependence of the variability.
On the
contrary, the flux variability on short time scales, with variations
by a factor of 2-3 on time scales of $\sim$ 4 ks, coupled to an
estimate of low values for the central black hole mass (even though
based more on assumptions that on measurements), could fit with
the observed trend of higher variability for smaller black hole 
masses, and could also be linked to variations in the accretion rate 
(O'Neill et al. 2005, Mushotzky et al. 1993).

Even if small, this sample has a statistically
sound definition that makes results applicable to other candidates.
We cannot estimate their space density, 
however we notice that several examples of objects with similar characteristics
have already appeared in the literature (Griffiths et al. 1996, Della Ceca et al. 2001, 
Guainazzi et al. 2000), which suggests that 
they might constitute a non-negligible fraction of the AGN family.

We therefore suggest the use of a combination of diagnostic ratios,
such as those based on the [OIII]/FIR ratio and the presence of X-ray emission 
to pinpoint members of this class of Composites.
New Spitzer Space Telescope and {\it Chandra} data
would be of crucial importance to deepen our knowledge on the 
Composite sample and to enlarge the number of candidates 
belonging to the same class.

\acknowledgments
FP warmly thanks for hospitality the Center for Astrophysics
where much of this work has been realized. We thank the
anonymous referee for providing useful comments. 
We acknowledge the contribution of Massimo Cappi and Mauro Dadina
for an earlier involvement in the paper,
Raffaella Landi for her help with the PDS data, Jonathan McDowell
for the SED-making program {\em TIGER}, 
Martin Elvis, and Andreas Zezas for useful comments.
This work has been supported by the NASA grants GO3-4131X and
NAS8-39073 to the {\it Chandra} X-Ray Center. This work has received
partial financial support by ASI and Cofin Miur.
RDC acknowledge partial financial support from the MIUR
(Cofin-03-02-23).
This research makes use of the NASA/IPAC Extra-galactic Database
(NED) and of data products from the Two Micron All
Sky Survey, which is a joint project of the University of
Massachusetts and the infrared Processing and Analysis
Center/California Institute of Technology, funded by the National
Aeronautics and Space Administration and the National Science
Foundation.

\clearpage

\begin{table*}[!ht]
	\begin{center}
     \caption{The Composite Seyfert/Star-forming sample}
 	\scriptsize
     \begin{tabular}{lccccccccc}
      \hline \hline
     {Name}  &RA &   DEC	& z &Hubble& N$_{\rm H}^{Gal}$ & Log L(FIR) & Log 
L$_{H\alpha}^{Tot}$  & Log L$_{H\alpha}^{Broad}$ & Log L$_{[O~III]}$ \\
             &(J2000)& (J2000) &   &Type& {$10^{20} cm^{-2}$}  &  & \\
      \hline
IRAS\,00317$-$2142 & 00:34:13.8 & $-$21:26:21 & 0.0268 &Sbc		& 1.55 &44.87&41.41 & 41.04  &41.11  \\
IRAS\,01072$+$4954 & 01:10:14.1 & $+$50:10:32 & 0.0237 &Sc		& 15.1 &44.09&40.64 & --   &40.70 \\
IRAS\,01319$-$1604 & 01:34:25.1 & $-$15:49:08 & 0.0199 &Sb 		& 1.42 &44.29&40.76 & --   &40.89 \\
IRAS\,04392$-$0123 & 04:41:48.2 & $-$01:18:06 & 0.0289 &Sc	        & 5.43 &44.26&40.71 & --   &41.01 \\
IRAS\,20051$-$1117 & 20:07:51.3 & $-$11:08:35 & 0.0315 &-		& 6.53 &44.88&41.01 & 40.67  &41.44 \\
IRAS\,20069$+$5929 & 20:07:50.8 & $+$59:38:10 & 0.0374 &Sc		& 12.7 &44.40&41.37 & 41.01  &41.52 \\
         \hline \hline							
     \end{tabular}							
   \label{sample}
	\end{center}
  \end{table*}

\clearpage

\begin{table*}[!ht]
	\begin{center}
     \caption{Chandra ACIS-I and {\it Beppo}SAX Target Observation details}
     \begin{tabular}{lccccc}
      \hline \hline
     {Name}  &   Obs. Date && Exp. (ks)& & Cts/s(0.3-10 keV) (10$^{-2}$) \\
\hline
             &   	  &&  ACIS-I  &  & ACIS-I \\
      \hline
IRAS\,01072+4954& 2003-09-17 && 22.19&  &5.34$\pm$0.28  \\
IRAS\,01319$-$1604& 2003-09-20 && 22.77&  &10.20$\pm$0.37 \\
IRAS\,04392$-$0123& 2003-03-25 && 23.58&  &1.43$\pm$0.10  \\
IRAS\,20069+5929& 2003-03-23 && 23.95&  &17.79$\pm$0.29 \\
         \hline  		
               &       	&  LECS   & MECS 	& PDS 	& (MECS) 	\\
      \hline
IRAS\,01072+4954 & 2002-01-02 & 11.65   & 21.00	&8.69	&0.32$\pm$0.06 \\
IRAS\,04392$-$0123 & 2001-09-14 & 26.32  & 57.38  &26.81  &0.26$\pm$0.03 \\
IRAS\,20051$-$1117 & 2001-10-29 & 10.80  & 56.06  &24.40  &3.02$\pm$0.17\\
       \hline \hline
   \label{infobs}
     \end{tabular}
\end{center}
\end{table*}

\begin{table*}
	\begin{center}
     \caption{Off-nuclear sources detected by {\it Chandra}.}
     \scriptsize
     \begin{tabular}{llllrrccc}
      \hline \hline
    {Galaxy}    && X-ray & Position     		&Offset   	&  Cts$\pm$Err &$\sigma$&F$_{2-10 keV}$  & Log L$_{2-10 keV}$\\
      {Name}   && RA & DEC    			&($^{\prime\prime}$/kpc)  	& & &   10$^{-15}$ cgs  &\\ 
      	
      \hline
IRAS\,01319$-$1604 &ulx-1 & 01:34:25.7 & $-$15:49:02.6 &9.7/3.9       & 12.7$\pm$4.0    &4.2    &3.8  &39.43\\
  	       &ulx-2 & 01:34:24.5 & $-$15:49:10.8 &10.6/4.3      & 9.6$\pm$3.5     &3.2    &2.9  &39.32\\
IRAS\,04392$-$0123 &ulx-1 & 04:41:47.6 & $-$01:17:43.9 &24.3/14.1     & 7.8$\pm$2.8     &3.9    &2.4  &39.56\\
%  	       &ulx-2 & 04:41:47.5 & $-$01:17:37.3 &31.2/18.1     & 4.7$\pm$2.2     &2.3    &1.4  &39.32\\
IRAS\,20069+5929 &ulx-1 & 20:07:51.4 & +59:38:11.8 &5.2/3.9       & 18.6$\pm$5.6    &4.0    &6.3  &40.20\\
         \hline \hline
   \label{off}
     \end{tabular}
\end{center}
\end{table*}

\clearpage

\clearpage

\begin{table*}
  	\begin{center}
%\begin{threeparttable}[b]
	    \caption{{\it Beppo}SAX spectral results.}
     \begin{tabular}{lcccccc}
          \hline \hline
     {Name}  & 	$N^{Gal}_{H}$ 	 & $N^{Int}_{H}$           & $\Gamma$& 
$\chi$$^{2}/dof$&F$_{2-10 keV}$ & L$_{2-10 keV}$ \\
             &{$10^{20} cm^{-2}$} &{$10^{22} cm^{-2}$}&         &		 
&10$^{-12}$ cgs  &Log		\\
      \hline
IRAS\,01072+4954 &  15.1  &$<$5.92 	& 1.87$^{+2.43}_{-0.85}$ & 2.3/8	& 
0.27 & 41.44\\
IRAS\,04392$-$0123 &  5.43  &$<$1.92	& 1.41$^{+0.80}_{-0.37}$ & 4/7		& 0.25 
& 41.57\\
IRAS\,20051$-$1117 &  6.53 &$<$0.12	&1.89$^{+0.09}_{-0.10}$  & 105/99	& 2.40 
& 42.63  \\
         \hline \hline	
   \label{bep}
             \end{tabular}	
  % \begin{tablenotes}
\tablecomments{Unabsorbed fluxes and intrinsic luminosities are
reported.}
  %  \end{tablenotes}
 %  \end{threeparttable}
	\end{center}	
  \end{table*}

\begin{table*}
  	\begin{center}
\scriptsize
	    \caption{Spectral results for the total sample from the combined {\it Chandra} and {\it Beppo}SAX data.}
     \begin{tabular}{lccccccccc}
          \hline \hline
     {Name}  & 		$N^{Gal}_{H}$ 	&  kT	& $N^{Int}_{H}$         & $\Gamma$ 
&EW& $\chi$$^{2}/dof$ &F$_{2-10 keV}$& L$_{2-10 keV}$\\
             &		{$10^{20} cm^{-2}$}& keV &{$10^{22} cm^{-2}$}&        & 
eV  &&10$^{-12}$ cgs &	Log	\\
      \hline
IRAS\,00317$-$2142\tablenotemark{a}	    &  1.55 & - 		     &0.08$^{+0.03}_{-0.03}$& 
1.91$^{+0.17}_{-0.15}$ & $<$1400& 47.3/49   & 0.27 &41.55\\
IRAS\,01072+4954 		    &  15.1 &  0.82$^{+0.17}_{-0.15}$ &     $<$0.04	 
  & 2.09$^{+0.13}_{-0.15}$ & $<$1170& 56.5/66   & 0.31 &41.49\\
IRAS\,01319$-$1604 		    &  1.42 &	    -		     &      $<$0.03	    & 
1.85$^{+0.06}_{-0.07}$ & $<$298 & 123.7/120 & 0.69 &41.69\\
IRAS\,04392$-$0123 		    &  5.43 &	    -		     &0.17$^{+0.2}_{-0.13}$ & 
1.34$^{+0.28}_{-0.21}$ &-       &  15.8/20  & 0.19 &41.46\\
IRAS\,20051$-$1117\tablenotemark{b}	    &  6.53 & - 		     &$<$0.02		    & 
1.79$^{+0.04}_{-0.04}$ &250$\pm$155&283.6/201&1.61 &42.46\\
IRAS\,20069+5929 		    &  12.7 &	    -		     &0.36$^{+0.06}_{-0.04}$& 
1.71$^{+0.08}_{-0.09}$ & $<$213 & 194.7/187 & 2.15 &42.74 \\
         \hline \hline	
   \label{spec}
     \end{tabular}	
     \vspace{-0.3cm}
\tablecomments{Unabsorbed fluxes and intrinsic luminosities are
reported.}
\tablenotetext{a}{Georgantopoulos, Zezas \& Ward (2003)}
\tablenotetext{b}{Georgantopoulos et al. (2004)}
 	\end{center}	
\end{table*}

\clearpage
\clearpage

\begin{table*}
	\begin{center}
     \caption{{\it Chandra} and {\it Beppo}SAX variability analysis.}
     \begin{tabular}{lcccccc}
      \hline \hline
     {Name}  	& cts/s    	& $\chi$$^{2}/dof$ & cts/s    	 
&$\chi$$^{2}/dof$ 	& cts/s    	&$\chi$$^{2}/dof$	\\
             	&   10$^{-2}$ 	&(0.3-10 keV)      &   10$^{-2}$ 	& (0.3-2 
keV)     	&   10$^{-2}$ 	&(2-10 keV)		\\
      \hline
IRAS\,01072+4954 		&5.34$\pm$0.28	  &210.2/25          &3.92$\pm$0.25	 
&177.0/24  	     &1.30$\pm$0.09   	&35.9/24	    	\\
IRAS\,01319$-$1604 		&10.20$\pm$0.37	  &265.3/25	     &7.34$\pm$0.32	 
&208.1/25	     &2.77$\pm$0.21	&60.1/25		\\
IRAS\,04392$-$0123 		&1.43$\pm$0.10     &19.2/25	     &0.78$\pm$0.08	 
&18.6/25	     &0.54$\pm$0.07	&12.6/22		\\
IRAS\,20051$-$1117\tablenotemark{a} &3.02$\pm$0.17     &29.1/18	 
&0.85$\pm$0.09  	   &21.7/18	     &2.13$\pm$0.14	&32.1/18	\\
IRAS\,20069+5929 		&17.79$\pm$0.29	  &46.8/25 	     &9.74$\pm$0.22	 
&33.9/25          &7.99$\pm$0.20   	&24.1/25 		\\
         \hline \hline
     \end{tabular}
\label{vartab}
\tablenotetext{a}{{\it Beppo}SAX data}
  	\end{center}
  \end{table*}

\clearpage

\clearpage

\begin{table*}
	\begin{center}
\scriptsize
     \caption{Long term variability in the 0.3-2 keV energy band}
     \begin{tabular}{lrlrlrlrll}
      \hline \hline
{Name}			&\multicolumn{2}{c}{RASS}  	& \multicolumn{2}{c}{ROSAT PSPC}&\multicolumn{2}{c}{{\it Beppo}SAX} &\multicolumn{2}{c}{\it Chandra}  & Ratio\\
         		&Date	&Flux	 	& Date	&Flux	    	& Date			&Flux       	& Date		&Flux        & Max/Min \\
      \hline
IRAS\,00317$-$2142 & Nov 1990& 5.18$\pm$0.52& Jun 1992  & 5.73$\pm$0.11	&Nov 1995     &0.79\tablenotemark{a} & Aug  2001 & 0.27\tablenotemark{b} &  21.2   \\
IRAS\,01072+4954   & Jan 1991& 1.01$\pm$0.44& -		&  -		&Jan 2002     &0.32$\pm$0.06 & Sept 2003 & 0.43$\pm$0.28 &   3.2   \\
IRAS\,01319$-$1604 & Nov 1990& 0.92$\pm$0.26& Jan 1992  & 1.98$\pm$0.17	&-     	      &-	     & Sept 2003 & 0.61$\pm$0.27 &   3.2   \\
IRAS\,04392$-$0123 & Jul 1990& 1.66$\pm$0.33& -		&  -		&Nov 2001     &0.09$\pm$0.009& Mar  2003 & 0.07$\pm$0.007&  23.7   \\
IRAS\,20051$-$1117 & Oct 1990& 1.63$\pm$0.33& -         &  -		&Oct 2001     &1.88$\pm$0.10 & Apr  2002 & 1.53\tablenotemark{c} &   1.2   \\
IRAS\,20069+5929   & Nov 1990& 0.74$\pm$0.24& -		&  -		&-	      &-	     & Mar  2003 & 1.33$\pm$0.03 &   1.8   \\
         \hline \hline	
     \end{tabular}	
   \label{long_tab}
\tablecomments{Unabsorbed fluxes are in units of 10$^{-12}$ erg cm$^{-2}$ s$^{-1}$,
ratios have been calculated using the highest and the lowest flux measured.}
\tablenotetext{a}{ASCA flux, Georgantopoulos (2000)} 
\tablenotetext{b}{Georgantopoulos, Zezas \& Ward (2003)}
 \tablenotetext{b}{Georgantopoulos et al. (2004)}  
  	\end{center}	
  \end{table*}

\clearpage

\begin{figure*}[htb]
\vbox{
\psfig{file=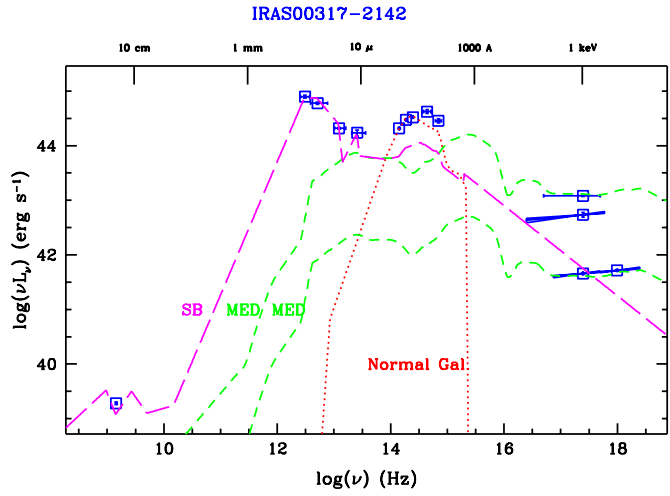,width=10truecm}\vfill
\psfig{file=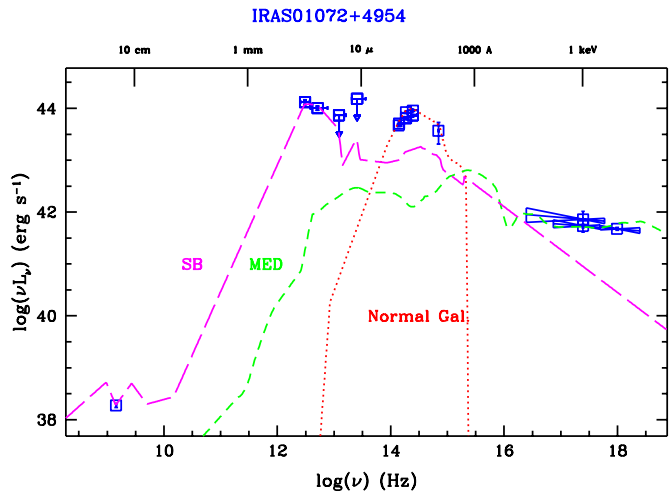,width=10truecm}
}
\vspace{-1in}
\caption{\scriptsize{Spectral Energy Distributions for the Composite objects.
Data points from the radio to the X-ray band have been plotted
with their relative errors. The {\it Chandra} spectral slope has
also been plotted. Each Composite SED have been
qualitatively compared with the Medium Energy Distribution (MED) 
template for radio 
quiet quasars (short dashed lines, Elvis et al. 1994), the
starburst galaxy template (long dashed lines, Schmitt et al. 1997) 
and the normal spiral galaxy template (dotted line, Elvis et al. 1994).
In the cases of IRAS\,00317$-$2142, IRAS\,01319$-$1604 and IRAS\,04392$-$0123,
the MED has been plotted twice to account for the strong 
X-ray flux variation.}}
\label{sed1}
\end{figure*}

\begin{figure*}[htb]
\vbox{
\psfig{file=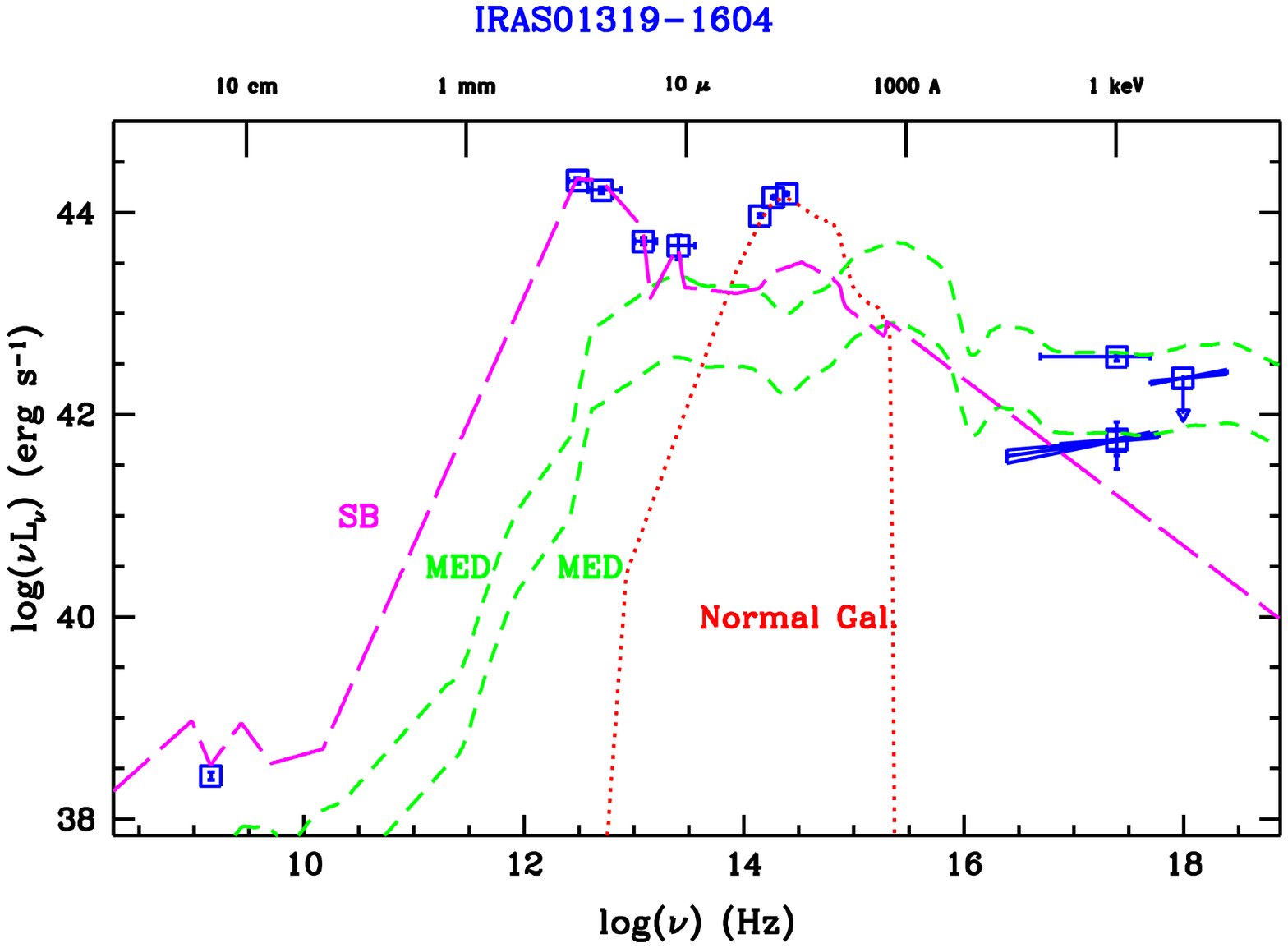,width=10truecm}\vfill
\psfig{file=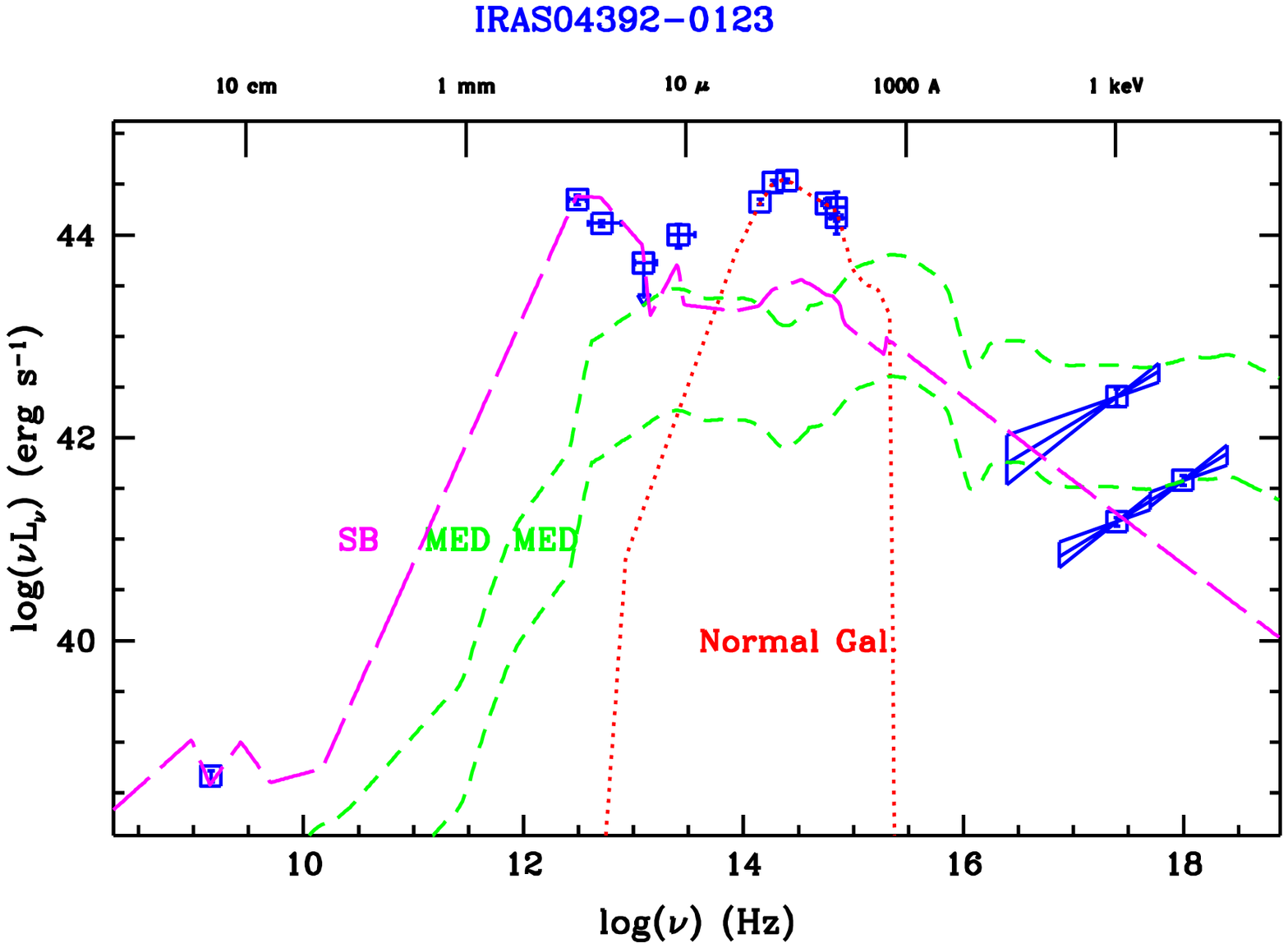,width=10truecm}
}
\vspace{-1in}
\setcounter{figure}{9}
\caption{{\it Continue}.}
\label{sed2}
\end{figure*}

\begin{figure*}[htb]
\vbox{
\psfig{file=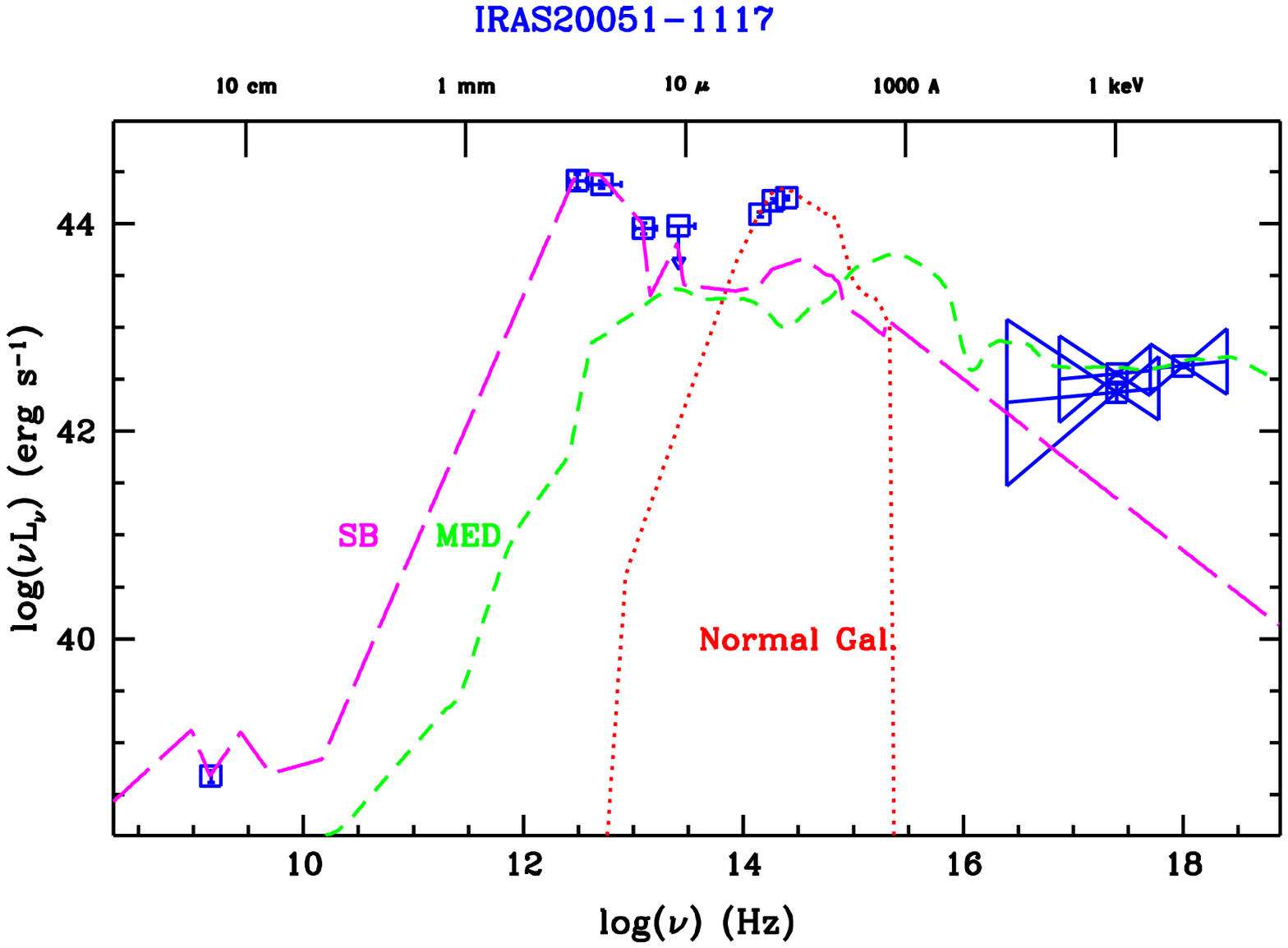,width=10truecm}\vfill
\psfig{file=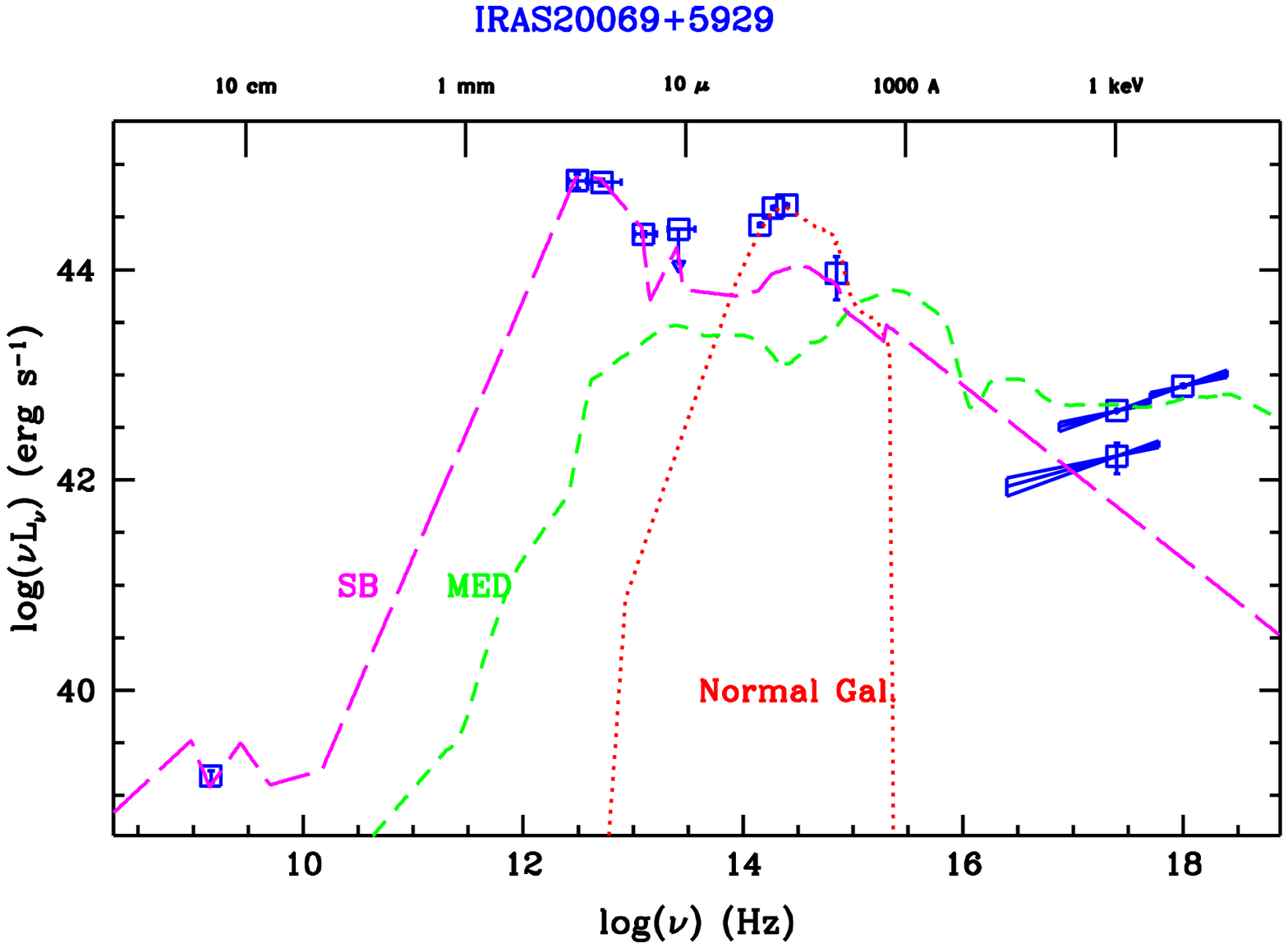,width=10truecm}
}
\vspace{-1in}
\setcounter{figure}{9}
\caption{{\it Continue}.}
\label{sed3}
\end{figure*}

\end{document}